\documentclass[10pt,a4paper]{article}

\usepackage[latin1]{inputenc} 
\usepackage[dvips]{graphicx} 
\usepackage{a4wide}
\usepackage{latexsym}
\usepackage{amsmath}
\usepackage{amssymb}
\usepackage{dsfont}

\title{Baryon vector and axial content up to the $7Q$ component}

\author{Cédric Lorcé\\ \small{\emph{Université de
Liège, Institut de Physique, Bât. B5a, B4000 Liège, Belgium}}\\
\small{\emph{Ruhr-Universität Bochum, Institut für Theoretische Physik II, D-44780 Bochum, Germany}}\\
\small{\emph{E-mail: C.Lorce@ulg.ac.be}}}
\date{}

\newcommand{\ud}{\mathrm{d}}
\newcommand{\uL}{\mathcal{L}}
\newcommand{\uM}{\mathcal{M}}

\newcommand{\uQcal}{\mathcal{Q}}
\newcommand{\uN}{\mathcal{N}}

\newcommand{\pslash}{p\!\!\!/}

\newcommand{\usigma}{\boldsymbol{\sigma}}

\newcommand{\upi}{\boldsymbol{\pi}}

\newcommand{\ux}{\mathbf{x}}

\newcommand{\ur}{\mathbf{r}}
\newcommand{\un}{\mathbf{n}}
\newcommand{\up}{\mathbf{p}}

\newcommand{\uq}{\mathbf{q}}
\newcommand{\uQ}{\mathbf{Q}}

\begin{document}

\maketitle

\begin{center}
\begin{minipage}[t]{15cm}
\small{We have used the light-cone formulation of Chiral-Quark
Soliton Model to investigate the vector and axial content of octet,
decuplet and hypothetical antidecuplet in the flavor $SU(3)$
symmetry limit. We have extended previous works by computing the
$7Q$ contribution to vector and axial charges for the octet and
antidecuplet but stayed at the $5Q$ sector for the decuplet where
the full computation needs much more time. As expected the $7Q$
component has a weaker impact on the quantities but still changes
them by a few percent. We give also a detailed decomposition of
those charges into flavor, valence quark, sea quark and antiquark
contributions. Many of them are of course not (yet) measured or
estimated and constitute then a theoretical estimation. Among the
different interesting observations made in this work are the
explicit quadrupole deformation of decuplet baryons due to the pion
field and the sum of quark spins larger than the pentaquark one.}
\end{minipage}
\end{center}

\section{Introduction}

Chiral-Quark Soliton Model ($\chi$QSM) has recently been formulated
on the light cone or, equivalently, in the Infinite Momentum Frame
(IMF) \cite{PetPol,DiaPet}. This provides a new approach for
extracting predictions out of the model. The light-cone formulation
is attractive in many ways. For example, light-cone wave functions
are particularly well suited to compute matrix elements of
operators. One can even choose to work in a specific frame where the
annoying part of currents, \emph{i.e.} pair creation and
annihilation part, does not contribute. On the top of that it is in
principle also easy to compute parton distributions once light-cone
wave functions are known.

The technique has already been used to study vector and axial
charges of the nucleon and $\Theta^+$ pentaquark width up to the
$5Q$ component without \cite{DiaPet} and with \cite{Moi} quark
orbital angular momentum. In this approach it has been shown that
relativistic corrections (quark angular momentum and sea-quark
pairs) reduce the naive quark model value $\frac{5}{3}$ for the
nucleon axial charge $g^{(3)}_A$ down to a value close to $1.257$
observed in beta decays.

The baryon structure is of capital importance for our understanding
of QCD. In this non-perturbative regime the theory cannot be solved
and models are needed to understand the physics at this scale. While
a picture of the baryon as a system of 3 nonrelativistic quarks
seems to explain rather well magnetic moments, masses and
meson-baryon couplings, one observes that in polarized Deep
Inelastic Scattering (DIS) processes there are other ingredients.
Let us mention for example the violation of Ellis-Jaffe sum rule
revealing the presence of hidden flavor in the nucleon. It has also
been observed that the quarks contribute only to $\sim 30\%$ of the
total nucleon spin leading to what is called the ``spin crisis''. It
is clear that the missing angular momentum can be attributed to
quark orbital momentum and gluon angular momentum. Unfortunately,
the individual contributions are not known. Many models try to
improve the so-called Naive Quark Model (NQM) by taking into account
other degrees of freedom and/or general features of QCD such as
special relativity and approximate chiral symmetry.

$\chi$QSM is a model based on chiral symmetry. A baryon is
considered as made of $N_C$ valence quarks living in a relativistic
mean chiral field. This mean field is a soliton with maximal
symmetry, namely a hedgehog pion field. A specific baryon then
corresponds to a specific rotational excitation of the solitonic
field. This model can be considered as some interpolation between
two \emph{a priori} orthogonal pictures: Constituent Quark Model
where baryons are made of valence quarks exclusively and Skyrme
Model where baryons are solitons of the pion field. $\chi$QSM has
both degrees of freedom. Here baryons are indeed made of valence
quarks but living in a solitonic relativistic mean chiral field. In
the limit where the pion field is weak, the Dirac sea is weakly
distorted and thus carry small energy $E_\textrm{sea}\simeq 0$. The
valence level is shallow $E_\textrm{lev}\simeq M_Q$ and hence the
valence quarks are nonrelativistic. This is very similar to the
Constituent Quark Model picture. In the limit where the pion field
is large, the bound-state level is so deep that it joins the Dirac
sea. The whole nucleon mass is given by $E_\textrm{sea}$ which can
be expanded in derivatives of the mean field, the first terms being
close to the Skyrme Model Lagrangian.

This model has been mostly studied in the so-called ``instant
form'', \emph{i.e.} with the usual parametrization of space-time
$x=(t,\ux)$ and reproduced successfully many experimental results
\cite{Baryon properties,TimGhil}. In the instant form, the sea can
be treated as a whole but a slowly rotating soliton approximation
has to be invoked. Although this approximation is well justified for
ordinary baryons (octet and decuplet) it is questionable for the
exotic ones (antidecuplet) \cite{DiaPet}. The light-cone approach to
$\chi$QSM is \emph{complementary}. Here we cannot treat the whole
Dirac sea at once. One has to perform an expansion of the baryon
wave function in Fock space. On the other hand, we can compute exact
rotations without referring to the large-$N_C$ limit for their
evaluation. Hence, there is \emph{a priori} no direct connection
between the moment of inertia of the soliton and the overlap of
individual quark wave function. Moreover, studying models on the
light cone is always very interesting since the description is
closer to experimental situation where baryons are usually moving
with high velocity.

In the IMF formulation of $\chi$QSM, it has been possible to write a
general expression for baryon light-cone wave functions. By
computing matrix elements of operators, one can access the flavor
and spin content of the baryons and work explicitly with
$0,1,2,\ldots ,n$ additional quark-antiquark pairs in a fully
relativistic way. On the top of that the solitonic approach allows
one to treat all light baryons in a simple and unique elegant way.
Since the $5Q$ component is important to understand the nucleon
structure one should by analogy care about the $7Q$ component in
pentaquark. On the top of that it is also an \emph{a posteriori}
check that the expansion in the number of quark-antiquark pairs is
justified.

This paper is organized as follows. In section \ref{Section deux} we
present in a short way the model formulated in the IMF and give the
explicit definitions of the quantities used. Then we indicate how to
compute the charges by means of matrix elements in each Fock sector
in section \ref{Section trois}. After contraction over all color,
spin, isospin and flavor indices one is left with scalar overlap
integrals. Physical quantities are then just specific linear
combinations of those scalar overlap integrals determined by $SU(3)$
symmetry. The explicit expressions of those integrals are presented
in section \ref{Section quatre}. Since our approach is restricted to
flavor $SU(3)$ symmetry we give tables making it explicit and
present the parametrization used in section \ref{Section cinq}. Our
results can be found in section \ref{Section six}. First we give the
formal combinations and then the numerical evaluation, followed by a
discussion and comparison with experimental knowledge.

\section{$\chi$QSM on the Light Cone}\label{Section deux}

Chiral-Quark Soliton Model ($\chi$QSM) is a model proposed to mimic
low-energy QCD. It emphasizes the role of constituent quarks of mass
$M$ and pseudoscalar mesons as the relevant degrees of freedom and
is based on the following effective Lagrangian
\begin{equation}
\uL_{\chi QSM}=\bar\psi(p)(\pslash-MU^{\gamma_5})\psi(p),
\end{equation}
where $U^{\gamma_5}$ is a (flavor) $SU(3)$ matrix. We used the
$SU(2)$ hedgehog \emph{Ansatz} for the soliton field trivially
embedded in $SU(3)$
\begin{equation}
U^{\gamma_5}=\left(\begin{array}{cc}
                     U_0 & 0 \\
                     0 & 1
                   \end{array}
\right),\qquad U_0=e^{i n^a\tau^a P(r)\gamma_5}
\end{equation}
with $\tau^a$ the usual $SU(2)$ Pauli matrices and $n^a=r^a/r$ the
unit vector pointing in the direction of $\ur$. Note that the
hedgehog \emph{Ansatz} implies that a rotation in ordinary space
($n^a$) can be compensated by a rotation in isospin space
($\tau^a$). The profile function $P(r)$ is determined by topological
constraints and minimization of the energy of the system.

Within this model it has been shown \cite{PetPol,DiaPet} that one
can write a general expression for $SU(3)$ baryon wave functions
\begin{equation}\label{Wavefunction}
|\Psi_B\rangle=\left[\prod_{\textrm{color}=1}^{N_C}\int(\ud\up)\,F(\up)\,a^\dag(\up)\right]\exp\left(\int(\ud\up)(\ud\up')\,a^\dag(\up)\,W(\up,\up')\,b^\dag(\up')\right)|\Omega_0\rangle.
\end{equation}
This expression may look somewhat complicated at first view but is
in fact really transparent. The model describes baryons as $N_C$
quarks populating the valence level with wave function $F$
accompanied by a whole sea of quark-antiquark pairs represented by
the coherent exponential. The wave function of such a
quark-antiquark pair is $W$. For a specific baryon, one has to
rotate each quark by a $SU(3)$-matrix $R$ and each antiquark by
$R^\dag$ and project the whole wave function on the quantum number
of the specific baryon $\int\ud R\, B_k^*(R)$, where $B_k^*(R)$
represents the way the baryon is transformed by $SU(3)$. The full
expression \cite{DiaPet} for the light-cone baryon wave function
contains color $\alpha$, flavor $f$, isospin $j$ and spin $\sigma$
indices
\begin{equation}
\begin{split}
|\Psi_k(B)\rangle&=\int\ud
R\,B_k^*(R)\,\epsilon^{\alpha_1\alpha_2\alpha_3}\left[\prod_{n=1}^3\int(\ud\up_n)\,R_{j_n}^{f_n}\,F^{j_n\sigma_n}(\up_n)\,a^\dag_{\alpha_n f_n \sigma_n}(\up_n)\right]\\
&\times\quad\exp\left(\int(\ud\up)(\ud\up')\,\delta_{\alpha'}^\alpha\,
a^\dag_{\alpha f\sigma}(\up)\,R^f_j\,
W^{j\sigma}_{j'\sigma'}(\up,\up')\,R^{\dag j'}_{f'}\,b^{\dag\alpha'
f'\sigma'}(\up') \right)|\Omega_0\rangle,\label{Full glory}
\end{split}
\end{equation}
where we have considered the physical case $N_C=3$. The three
valence quarks are always antisymmetric in color
$\epsilon^{\alpha_1\alpha_2\alpha_3}$ and the additional
quark-antiquark pairs are color singlets $\delta_{\alpha'}^\alpha$.
This wave function is supposed provide a lot of information about
all light baryons.

\subsection{Valence wave function}
\begin{figure}[h!]\begin{center}\begin{minipage}[c]{8cm}\begin{center}\includegraphics[width=8cm]{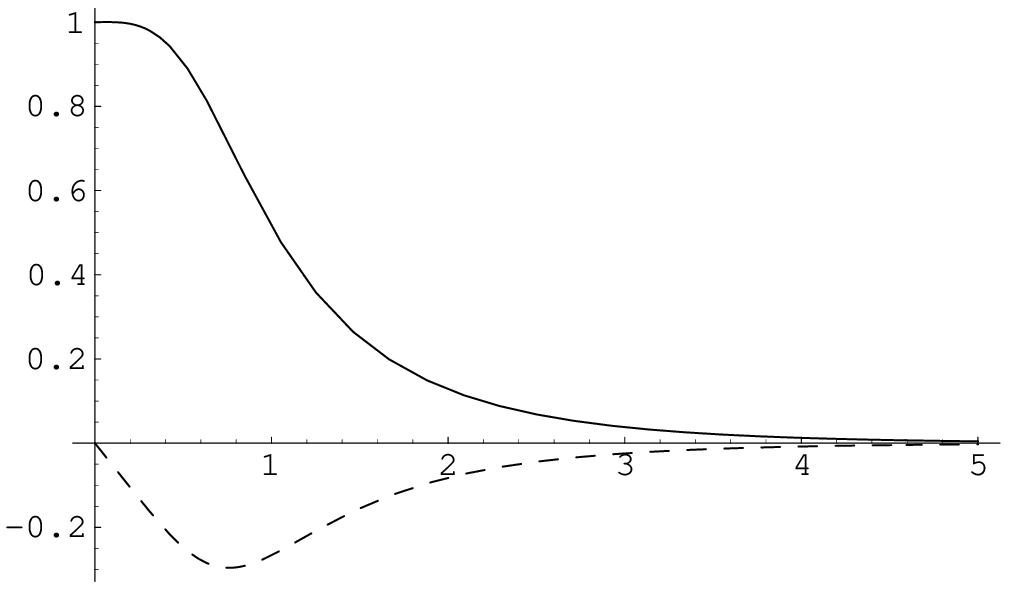}\caption{\small{Upper $s$-wave component $h(r)$ (solid) and lower $p$-wave component
$j(r)$ (dashed) of the bound-state quark level in light baryons.
Each of the three valence quarks has energy $E_\textrm{lev}=200$
MeV. Horizontal axis has units of $1/M=0.57$ fm.}}\label{Level}
\end{center}\end{minipage}\hspace{0.5cm}
\begin{minipage}[c]{8cm}\begin{center}\includegraphics[width=8cm]{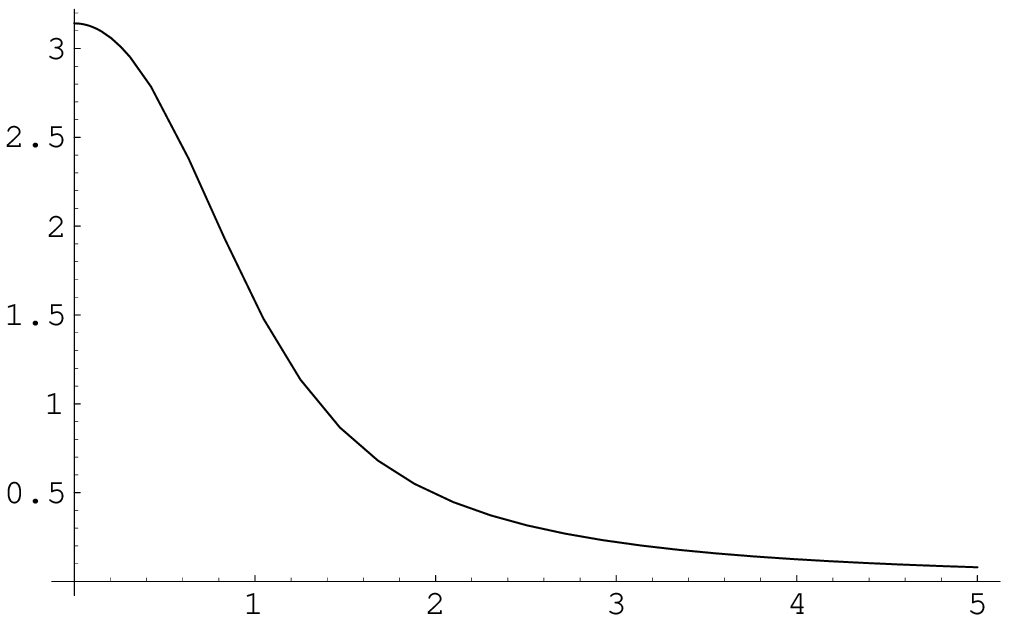}\caption{\small{Profile of
the self-consistent chiral field $P(r)$ in light baryons. The
horizontal axis unit is $r_0=0.8/M=0.46$
fm.\newline\newline\newline}}\label{Profile}
\end{center}\end{minipage}\end{center}
\end{figure}

On the light cone the valence level wave function $F$ is given by
\begin{equation}\label{Discrete level IMF}
F^{j\sigma}_\textrm{lev}(z,\up_\perp)=\sqrt{\frac{\uM}{2\pi}}\left[\epsilon^{j\sigma}h(p)+(p_z\mathds{1}+i\up_\perp\times\tau_\perp)^\sigma_{\sigma'}\epsilon^{j\sigma'}\frac{j(p)}{|\up|}\right]_{p_z=z\uM-E_\textrm{lev}}
\end{equation}
where $j$ and $\sigma$ are isospin\footnote{We remind that due to
the hedgehog \emph{Ansatz} rotations in ordinary space are
equivalent to isospin rotations. That is the reason why $j$ has been
called isospin index even though it can be seen as total angular
momentum of the quark.} and spin indices respectively, $z$ is the
fraction of baryon longitudinal momentum carried by the quark,
$\up_\perp$ is its transverse momentum and $\uM$ is the classical
soliton mass. The functions $h(p)$ and $j(p)$ are Fourier transforms
of the upper ($L=0$) $h(r)$ and lower ($L=1$) $j(r)$ components of
the spinor solution (see Fig.\ref{Level}) of the static Dirac
equation in the mean field with eigenenergy\footnote{This
eigenenergy turned out to be $E_\textrm{lev}\approx 200$ MeV when
solving the system of equations self-consistently for constituent
quark mass $M=345$ MeV.} $E_\textrm{lev}$
\begin{equation}
\psi_\textrm{lev}(\ux)=\left(\begin{array}{c}\epsilon^{ji}h(r)\\-i\epsilon^{jk}(\un\cdot\sigma)^i_k\,
j(r)\end{array}\right),\qquad\left\{\begin{array}{c}h'+h\,M\sin
P-j(M\cos P+E_\textrm{lev})=0\\j'+2j/r-j\,M\sin P-h(M\cos
P-E_\textrm{lev})=0\end{array}\right. ,
\end{equation}
where $P(r)$, the profile function of the soliton, is fairly
approximated by \cite{Profile function, Approximation} (see
Fig.\ref{Profile})
\begin{equation}\label{Profile function}
P(r)=2\arctan\left(\frac{r_0^2}{r^2}\right),\qquad
r_0\approx\frac{0.8}{M}.
\end{equation}
\subsection{Pair wave function}
The quark-antiquark pair wave function $W$ can be written in terms
of the Fourier transform of the chiral field with chiral circle
condition $\Pi^2+\Sigma^2=1$, $U_0=\Sigma+i\Pi\gamma_5$. The chiral
field is then given by
\begin{equation} \Pi=\un\cdot\tau\sin P(r),\qquad \Sigma(r)=\cos P(r)
\end{equation}
and its Fourier transform by
\begin{equation}\label{Fourier tranform of mean field}
\Pi(\uq)^j_{j'}=\int\ud^3\ux\,
e^{-i\uq\cdot\ux}(\un\cdot\tau)^j_{j'}\sin
P(r),\qquad\Sigma(\uq)^j_{j'}=\int\ud^3\ux\, e^{-i\uq\cdot\ux}(\cos
P(r)-1)\delta^j_{j'},
\end{equation}
where $j$ and $j'$ are the isospin indices of the quark and
antiquark, respectively. The pair wave function is obtained by
considering the expansion of the quark propagator \cite{PetPol} in
the mean field in terms of the chiral interaction $V=U_0-1$. After
the boost to the IMF, the pair wave function appears as a function
of the fractions of the baryon longitudinal momentum carried by the
quark $z$ and antiquark $z'$ of the pair and their transverse
momenta $\up_\perp$, $\up'_\perp$
\begin{equation}
W^{j\sigma}_{j'\sigma'}(z,\up_\perp;z',\up'_\perp)=\frac{M\uM}{2\pi
Z}\left\{\Sigma^j_{j'}(\uq)[M(z'-z)\tau_3+\uQ_\perp\cdot\tau_\perp]^\sigma_{\sigma'}+i\Pi^j_{j'}(\uq)[-M(z'+z)\mathds{1}+i\uQ_\perp\times\tau_\perp]^\sigma_{\sigma'}\right\},
\end{equation}
where $\uq=((\up+\up')_\perp,(z+z')\uM)$ is the three-momentum of
the pair as a whole transferred from the background fields
$\Sigma(\uq)$ and $\Pi(\uq)$. As earlier $j$ and $j'$ are isospin
and $\sigma$ and $\sigma'$ are spin indices with the prime for the
antiquark. In order to condense the notations we used
\begin{equation}\label{Notation}
Z=\uM^2zz'(z+z')+z(p'^2_\perp+M^2)+z'(p^2_\perp+M^2),\qquad
\uQ_\perp=z\up'_\perp-z'\up_\perp.
\end{equation}
A more compact form for this wave function can be obtained by means
of the following two variables
\begin{equation}
y=\frac{z'}{z+z'},\qquad\uQcal_\perp=\frac{z\up'_\perp-z'\up_\perp}{z+z'}.
\end{equation}
The pair wave function then takes the form
\begin{equation}\label{Pair wavefunction}
W^{j\sigma}_{j'\sigma'}(y,\uq,\uQcal_\perp)=\frac{M\uM}{2\pi
}\,\frac{\Sigma^j_{j'}(\uq)[M(2y-1)\tau_3+\uQcal_\perp\cdot\tau_\perp]^\sigma_{\sigma'}+i\Pi^j_{j'}(\uq)[-M\mathds{1}+i\uQcal_\perp\times\tau_\perp]^\sigma_{\sigma'}}{\uQcal^2_\perp+M^2+y(1-y)\uq^2}.
\end{equation}

\subsection{Rotational wave function}\label{Projection}
To obtain the wave function of a specific baryon with given spin
projection $k$, one has to rotate the soliton in ordinary and flavor
spaces and then project on quantum numbers of this specific baryon.
For example, one has to compute the following integral to obtain the
neutron rotational wave function in the $3Q$ sector
\begin{equation}
T(n^0)_{k,j_1j_2j_3}^{f_1f_2f_3}=\int\ud R\,n_k(R)^*
R^{f_1}_{j_1}R^{f_2}_{j_2}R^{f_3}_{j_3},
\end{equation}
where $R$ is a $SU(3)$ matrix and
$n_k(R)^*=\frac{\sqrt{8}}{24}\,\epsilon_{kl}R^{\dag l}_2R^3_3$
represents the way that the neutron is transformed under $SU(3)$
rotations. This integral means that the neutron state $n_k(R)^*$ is
projected onto the $3Q$ sector
$R^{f_1}_{j_1}R^{f_2}_{j_2}R^{f_3}_{j_3}$ by means of the
integration over all $SU(3)$ matrices $\int\ud R$. By contracting
this rotational wave function $T(n^0)_{k,j_1j_2j_3}^{f_1f_2f_3}$
with the nonrelativistic $3Q$ wave function\footnote{The
nonrelativistic limit here means that we neglect the lower component
$j$ of the Dirac field.}
$\epsilon^{j_1\sigma_1}\epsilon^{j_2\sigma_2}\epsilon^{j_3\sigma_3}h(p_1)h(p_2)h(p_3)$
one finally obtains the nonrelativistic neutron wave function
\begin{equation}
|n^0\rangle_k^{f_1f_2f_3,\sigma_1\sigma_2\sigma_3}=\frac{\sqrt{8}}{24}\,\epsilon^{f_1f_2}\epsilon^{\sigma_1\sigma_2}\delta^{f_3}_2\delta^{\sigma_3}_kh(p_1)h(p_2)h(p_3)+\textrm{cyclic
permutations of 1,2,3.}
\end{equation}
This expression means\footnote{One has $f=u,d,s$ and
$\sigma=\uparrow,\downarrow$.} that there is a $ud$ pair in
spin-isospin zero combination
$\epsilon^{f_1f_2}\epsilon^{\sigma_1\sigma_2}$ and that the third
quark is a down quark $\delta^{f_3}_2$ and carries the whole spin of
the neutron $\delta^{\sigma_3}_k$. This is in fact exactly the
$SU(6)$ spin-flavor wave function for the neutron.

The rotational wave function of octet, decuplet and antidecuplet in
the $3Q$, $5Q$ and $7Q$ sectors can all be found in the Appendix of
this paper.

\section{Currents, charges and matrix elements}\label{Section trois}

A typical physical observable is the matrix element of some operator
(preferably written in terms of quark annihilation-creation
operators $a$, $b$, $a^\dag$, $b^\dag$) sandwiched between the
initial and final baryon wave functions. These wave functions are
superpositions of Fock states obtained by expanding the coherent
exponential in eq. \eqref{Wavefunction}. One can reasonably expect
that the Fock states with the lowest number of quarks will give the
main contribution. If one uses the Drell frame $q^+=0$
\cite{Drell,Brodsky} where $q$ is the total momentum transfer, then
the vector $\bar\psi\gamma^+\psi$ and axial
$\bar\psi\gamma^+\gamma^5\psi$ currents can neither create nor
annihilate any quark-antiquark pair. This is a big advantage of the
light-cone formulation since one needs to compute diagonal
transitions only, \emph{i.e.} $3Q$ into $3Q$, $5Q$ into $5Q$,
$\ldots$ and not $3Q$ into $5Q$ for example.

In the $3Q$ sector, since all (valence) quarks are on the same
footing, all the possible contractions of creation-annihilation
operators are equivalent. One can use a diagram to represent these
contractions. The contractions without any current operator acting
on a quark line correspond to the normalization of the state. We
choose the simplest one where all quarks with the same label are
connected, see Fig.\ref{Three-q normalization}.
\begin{figure}[h]\begin{center}\begin{minipage}[c]{8cm}\begin{center}\includegraphics[width=4cm]{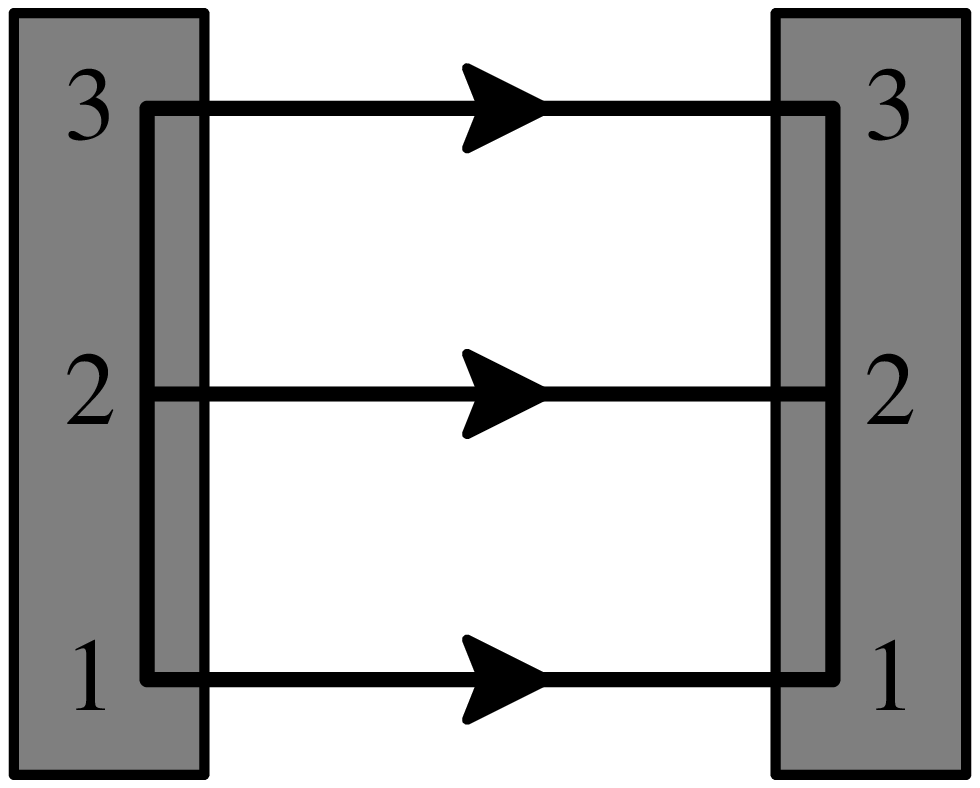}
\caption{\small{Schematic representation of the $3Q$ normalization.
Each quark line stands for the color, flavor and spin contractions
$\delta^{\alpha_i}_{\alpha'_i}\delta^{f_i}_{f'_i}\delta^{\sigma_i}_{\sigma'_i}
\int\ud z'_i\,\ud^2\up'_{i\perp}\delta(z_i-z'_i)$
$\delta^{(2)}(\up_{i\perp}-\up'_{i\perp})$. The large dark
rectangles stand for the three initial (left) and final (right)
valence quarks antisymmetrized in color
$\epsilon_{\alpha_1\alpha_2\alpha_3}$.}}\label{Three-q
normalization}
\end{center}\end{minipage}\hspace{0.5cm}
\begin{minipage}[c]{8cm}\begin{center}\includegraphics[width=3.5cm]{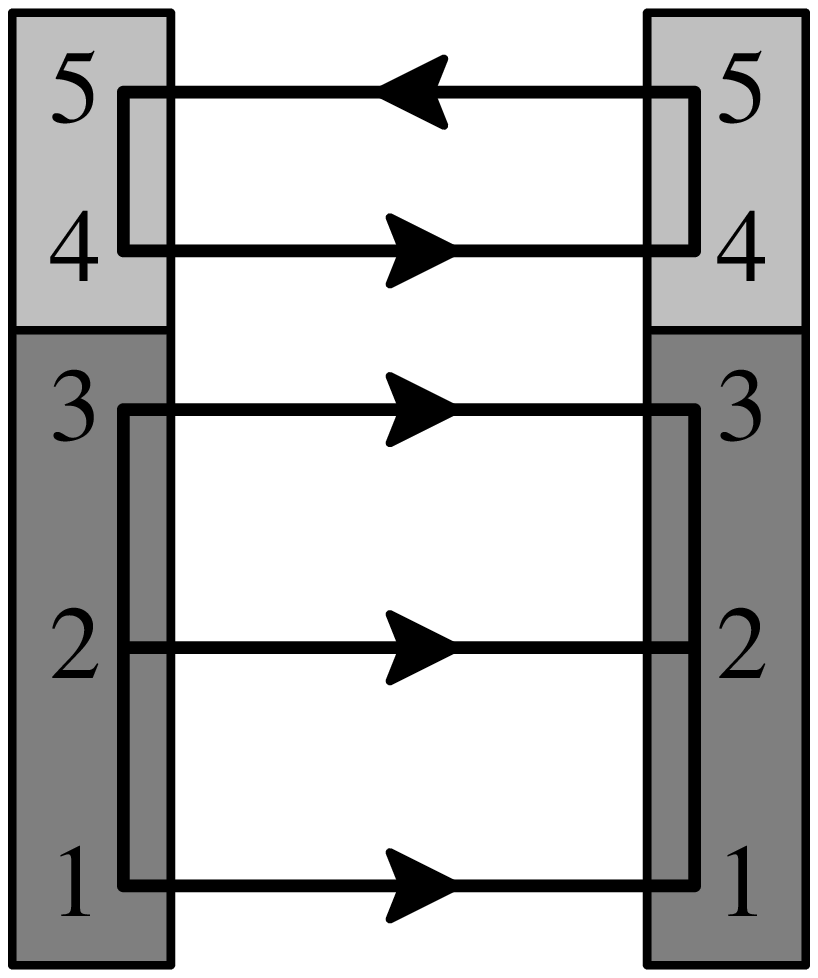}\hspace{1cm}\includegraphics[width=3.5cm]{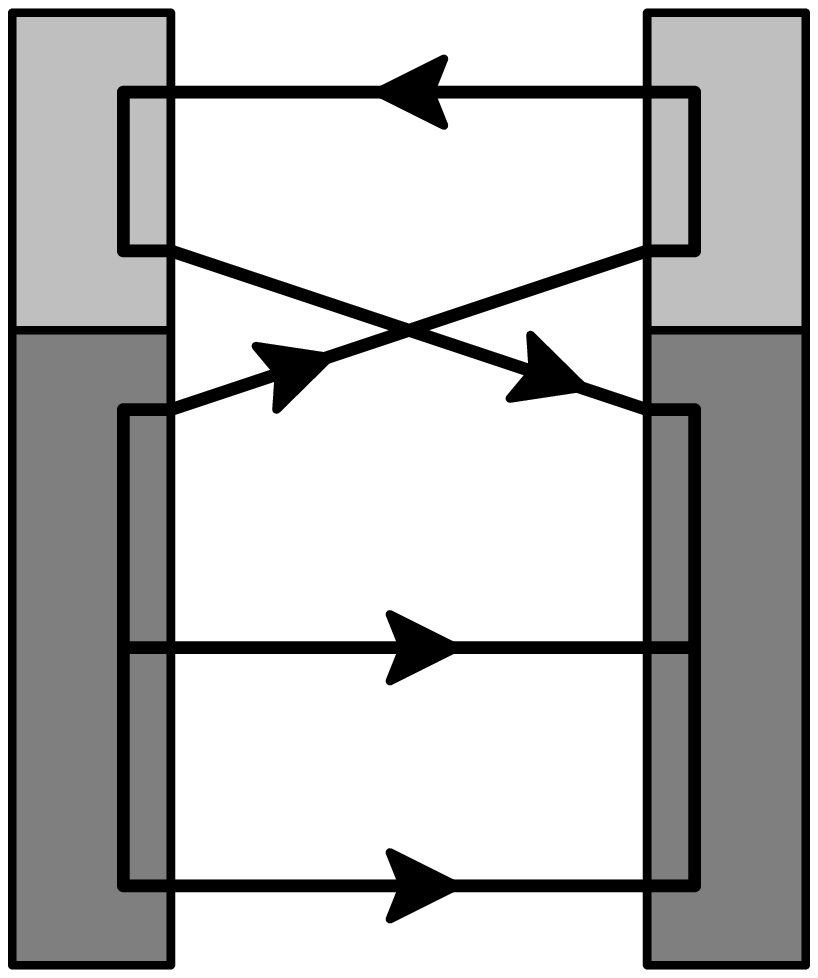}
\caption{\small{Schematic representation of the $5Q$ direct (left)
and exchange (right) contributions to the normalization. The
quark-antiquark pairs are represented by small light rectangles and
are in color singlet $\delta^{\alpha_4}_{\alpha_5}$.}}\label{Five-q
normalization}
\end{center}\end{minipage}\end{center}
\end{figure}

In the $5Q$ sector, all contractions are equivalent to either the
so-called ``direct'' diagram or the ``exchange'' diagram, see
Fig.\ref{Five-q normalization}. In the direct diagram, all quarks
with the same label are connected while in the exchange one, a
valence quark is exchanged with the quark of the sea pair. It has
appeared in a previous work \cite{Moi} that exchange diagrams do not
contribute much and can thus be neglected (there is no disconnected
quark loop). So we use only the direct contributions throughout this
paper.

\begin{figure}[h]\begin{center}\includegraphics[width=15cm]{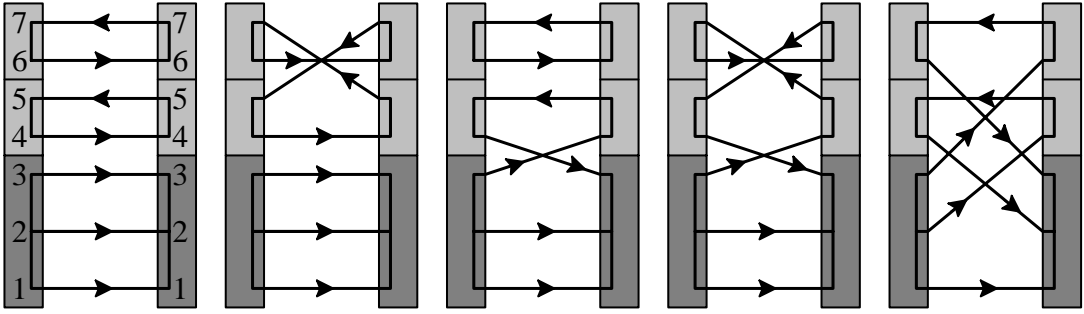}
\caption{\small{Schematic representation of the $7Q$ contributions
to the normalization.}}\label{Sevenquarks}\end{center}
\end{figure}

In the $7Q$ sector there are 5 types of diagrams, see
Fig.\ref{Sevenquarks}. The three last diagrams involve at least an
exchange of a valence quark with a sea quark. Those are neglected in
the present work by analogy with the $5Q$ sector. In the second and
fourth diagrams the two pairs exchange their quark (or antiquark)
and are likely negligible. We therefore expect that the first
diagram gives the major contribution in the $7Q$ sector. A
mathematical argument is that contraction over color indices favors
this diagram by at least a factor 3 (there is at least one more
disconnected quark loop compared to the other diagrams). A physical
argument would be that this diagram represents a process where
nothing really happens and is thus expected to be dominant compared
to the other diagrams where quarks exchange their roles.

The vector and axial operators act on each quark line. In the
present approach it is easy to compute separately the contributions
coming from the valence quarks, the sea quarks and antiquarks, see
Fig.\ref{Direct charges}.
\begin{figure}[h]\begin{center}\includegraphics[width=10cm]{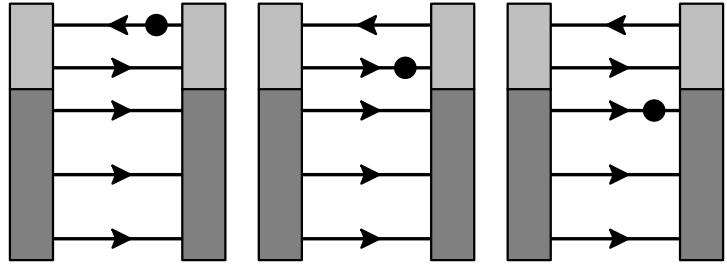}
\caption{\small{Schematic representation of the three types of $5Q$
contributions to the charges: antiquark (left), sea quark (center)
and valence quark (right) contributions.}}\label{Direct
charges}\end{center}
\end{figure}
These diagrams represent some contraction of color, spin, isospin
and flavor indices. For example, the sum of the three diagrams in
the $5Q$ sector with the vector current acting on the quark lines
represents the following expression
\begin{equation}
\begin{split}
V^{(5)}(1\to 2)&=\frac{108}{2}\,\delta^k_l\,T(1)^{f_1f_2f_3f_4,j_5}_{j_1j_2j_3j_4,f_5,k}\,T(2)_{f_1f_2g_3g_4,l_5}^{l_1l_2l_3l_4,g_5,l}\int(\ud p_{1-5})\\
&\times F^{j_1\sigma_1}(p_1)F^{j_2\sigma_2}(p_2)F^{j_3\sigma_3}(p_3)W^{j_4\sigma_4}_{j_5\sigma_5}(p_4,p_5)F^\dag_{l_1\sigma_1}(p_1)F^\dag_{l_2\sigma_2}(p_2)F^\dag_{l_3\tau_3}(p_3)W_{c\,l_4\tau_4}^{l_5\tau_5}(p_4,p_5)\\
&\times \left[-\delta^{g_3}_{f_3}\delta^{g_4}_{f_4}\boldsymbol{
J^{f_5}_{g_5}}\delta^{\tau_3}_{\sigma_3}\delta^{\tau_4}_{\sigma_4}\boldsymbol{\delta_{\tau_5}^{\sigma_5}}+\delta^{g_3}_{f_3}\boldsymbol{
J^{g_4}_{f_4}}\delta^{f_5}_{g_5}\delta^{\tau_3}_{\sigma_3}\boldsymbol{\delta^{\tau_4}_{\sigma_4}}\delta_{\tau_5}^{\sigma_5}+3\boldsymbol{
J^{g_3}_{f_3}}\delta^{g_4}_{f_4}\delta^{f_5}_{g_5}\boldsymbol{\delta^{\tau_3}_{\sigma_3}}\delta^{\tau_4}_{\sigma_4}\delta_{\tau_5}^{\sigma_5}\right],\label{Direct}
\end{split}
\end{equation}
where $J^f_g$ is the flavor content of the current. The axial charge
is easily obtained from the vector one. One just has to replace the
averaging over baryon spin by $\frac{1}{2}(-\sigma_3)^k_l$ and the
axial charge operator involves now
$\boldsymbol{(-\sigma_3)^{\tau_i}_{\sigma_i}}$ instead of
$\boldsymbol{\delta^{\tau_i}_{\sigma_i}}$. One then has
\begin{equation}
\begin{split}
A^{(5)}(1\to 2)&=\frac{108}{2}\,(-\sigma_3)^k_l\,T(1)^{f_1f_2f_3f_4,j_5}_{j_1j_2j_3j_4,f_5,k}\,T(2)_{f_1f_2g_3g_4,l_5}^{l_1l_2l_3l_4,g_5,l}\int(\ud p_{1-5})\\
&\times F^{j_1\sigma_1}(p_1)F^{j_2\sigma_2}(p_2)F^{j_3\sigma_3}(p_3)W^{j_4\sigma_4}_{j_5\sigma_5}(p_4,p_5)F^\dag_{l_1\sigma_1}(p_1)F^\dag_{l_2\sigma_2}(p_2)F^\dag_{l_3\tau_3}(p_3)W_{c\,l_4\tau_4}^{l_5\tau_5}(p_4,p_5)\\
&\times \left[-\delta^{g_3}_{f_3}\delta^{g_4}_{f_4}\boldsymbol{
J^{f_5}_{g_5}}\delta^{\tau_3}_{\sigma_3}\delta^{\tau_4}_{\sigma_4}\boldsymbol{(-\sigma_3)_{\tau_5}^{\sigma_5}}+\delta^{g_3}_{f_3}\boldsymbol{
J^{g_4}_{f_4}}\delta^{f_5}_{g_5}\delta^{\tau_3}_{\sigma_3}\boldsymbol{(-\sigma_3)^{\tau_4}_{\sigma_4}}\delta_{\tau_5}^{\sigma_5}+3\boldsymbol{
J^{g_3}_{f_3}}\delta^{g_4}_{f_4}\delta^{f_5}_{g_5}\boldsymbol{(-\sigma_3)^{\tau_3}_{\sigma_3}}\delta^{\tau_4}_{\sigma_4}\delta_{\tau_5}^{\sigma_5}\right].
\label{Direct}
\end{split}
\end{equation}

\section{Scalar overlap integrals}\label{Section quatre}

The contractions in previous section are easily performed by
\emph{Mathematica} over all flavor $(f,g)$, isospin $(j,l)$ and spin
$(\sigma,\tau)$ indices. One is then left with scalar integrals over
longitudinal $z$ and transverse $\up_\perp$ momenta of the quarks.
The integrals over relative transverse momenta in the
quark-antiquark pair are generally UV divergent. We have chosen to
use the Pauli-Villars regularization with mass $M_\textrm{PV}=556.8$
MeV (this value being chosen from the requirement that the pion
decay constant $F_\pi=93$ MeV is reproduced for $M=345$ MeV).

For convenience we introduce the probability distribution
$\Phi^I(z,\uq_\perp)$ seen by a vector ($I=V$) or an axial ($I=A$)
probe, that three valence quarks leave the longitudinal fraction
$z=q_z/\uM$ and the transverse momentum $\uq_\perp$ to the
quark-antiquark pair(s)
\begin{equation}\label{Probability3q}
\Phi^I(z,\uq_\perp)=\int\ud
z_{1,2,3}\frac{\ud^2\up_{1,2,3\perp}}{(2\pi)^6}\,\delta(z+z_1+z_2+z_3-1)(2\pi)^2\delta^{(2)}(\uq_\perp+\up_{1\perp}+\up_{2\perp}+\up_{3\perp})D^I(p_1,p_2,p_3).
\end{equation}
The function $D^I(p_1,p_2,p_3)$ is given in terms of the upper and
lower valence wave functions $h(p)$ and $j(p)$ as follows
\begin{equation}
\begin{split}
D^V(p_1,p_2,p_3)&=h^2_1h^2_2h^2_3+6h^2_1h^2_2\left[h_3\frac{p_{3z}}{|\up_3|}j_3\right]+3h^2_1h^2_2j^2_3+12h^2_1\left[h_2\frac{p_{2z}}{|\up_2|}j_2\right]\left[h_3\frac{p_{3z}}{|\up_3|}j_3\right]\\
&+12h^2_1\left[h_2\frac{p_{2z}}{|\up_2|}j_2\right]j^2_3+8\left[h_1\frac{p_{1z}}{|\up_1|}j_1\right]\left[h_2\frac{p_{2z}}{|\up_2|}j_2\right]\left[h_3\frac{p_{3z}}{|\up_3|}j_3\right]+3h^2_1j^2_2j^2_3\\
&+12\left[h_1\frac{p_{1z}}{|\up_1|}j_1\right]\left[h_2\frac{p_{2z}}{|\up_2|}j_2\right]j^2_3+6\left[h_1\frac{p_{1z}}{|\up_1|}j_1\right]j^2_2j^2_3+j^2_1j^2_2j^2_3,\label{Phi}
\end{split}
\end{equation}
\begin{equation}
\begin{split}
D^A(p_1,p_2,p_3)&=h^2_1h^2_2h^2_3+6h^2_1h^2_2\left[h_3\frac{p_{3z}}{|\up_3|}j_3\right]+h^2_1h^2_2\frac{2p_{3z}^2+p_3^2}{p_3^2}j^2_3+12h^2_1\left[h_2\frac{p_{2z}}{|\up_2|}j_2\right]\left[h_3\frac{p_{3z}}{|\up_3|}j_3\right]\\
&+4h^2_1\left[h_2\frac{p_{2z}}{|\up_2|}j_2\right]\frac{2p_{3z}^2+p_3^2}{p_3^2}j^2_3+8\left[h_1\frac{p_{1z}}{|\up_1|}j_1\right]\left[h_2\frac{p_{2z}}{|\up_2|}j_2\right]\left[h_3\frac{p_{3z}}{|\up_3|}j_3\right]+h^2_1j^2_2\frac{4p_{3z}^2-p_3^2}{p_3^2}j^2_3\\
&+4\left[h_1\frac{p_{1z}}{|\up_1|}j_1\right]\left[h_2\frac{p_{2z}}{|\up_2|}j_2\right]\frac{2p_{3z}^2+p_3^2}{p_3^2}j^2_3+2\left[h_1\frac{p_{1z}}{|\up_1|}j_1\right]j^2_2\frac{4p_{3z}^2-p_3^2}{p_3^2}j^2_3+j^2_1j^2_2\frac{2p_{3z}^2-p_3^2}{p_3^2}j^2_3,\label{Psi}
\end{split}
\end{equation}
where we have used $h_i\equiv h(p_i)$ and $j_i\equiv j(p_i)$.

In the nonrelativistic limit one has $j(p)=0$ and thus
$D^V(p_1,p_2,p_3)=D^A(p_1,p_2,p_3)$ as it should be. Indeed,
nonrelativistic quarks have no orbital angular momentum and then
axial and vector probes see the same valence quark distribution. In
other words, because of the absence of quark angular momentum, a
quark with helicity $\pm$ has spin $z$-projection $\pm 1/2$,
respectively.

\subsection{$3Q$ scalar integrals}

In the $3Q$ sector there is no quark-antiquark pair. There are then
two integrals only, one for the vector case
\begin{equation}
\Phi^V(0,0)
\end{equation}
and one for the axial one
\begin{equation}
\Phi^A(0,0),
\end{equation}
where the null argument indicates that the whole baryon momentum is
carried by the three valence quarks. Let us remind that in this
sector, spin-flavor wave functions obtained by the projection
technique are equivalent to those given by $SU(6)$ symmetry. One
then naturally obtains the same results for the charges as those
given by $SU(6)$ NQM, excepted that axial quantities are multiplied
by the factor $\Phi^A(0,0)/\Phi^V(0,0)$. This is similar to the
usual approach based on the Melosh rotation \cite{Melosh}. In usual
light-cone models one starts with nonrelativistic $SU(6)$ wave
functions and then performs a Melosh rotation on the spinors to
obtain the helicity basis, particularly well suited for light-cone
treatment. This rotation introduces orbital angular momentum
somewhat artificially. The net effect of this rotation is the
introduction of a Melosh factor to the observables compared with NQM
predictions
\begin{equation}
q=M_V\,q_{NQM},\qquad \Delta q=M_A\,\Delta q_{NQM}.
\end{equation}
We will discuss this point more intensively in a further work.

\subsection{$5Q$ scalar integrals}

In the $5Q$ sector there is one quark-antiquark pair and only seven
integrals are needed. These integrals can be written in the general
form
\begin{equation}
K^I_J=\frac{M^2}{2\pi}\int \frac{\ud^3
\uq}{(2\pi)^3}\,\Phi^I\left(\frac{q_z}{\uM},\uq_\perp\right)\theta(q_z)\,q_z\,G_J(q_z,\uq_\perp),
\end{equation}
where $G_J$ is a quark-antiquark probability distribution and
$J=\pi\pi,33,\sigma\sigma,3\sigma$. These distributions are obtained
by contracting two quark-antiquark wave functions $W$, see eq.
\eqref{Pair wavefunction} and regularized by means of Pauli-Villars
procedure
\begin{subequations}\label{Direct}
\begin{align}
G_{\pi\pi}(q_z,\uq_\perp)&=\Pi^2(\uq)\int^1_0\ud
y\int\frac{\ud^2\uQcal_\perp}{(2\pi)^2}\left[\frac{\uQcal^2_\perp+M^2}{(\uQcal^2_\perp+M^2+y(1-y)\uq^2)^2}-(M\to
M_\textrm{PV})\right],
\\
G_{33}(q_z,\uq_\perp)&=\frac{q_z^2}{\uq^2}\,G_{\pi\pi}(q_z,\uq_\perp),
\\
G_{\sigma\sigma}(q_z,\uq_\perp)&=\Sigma^2(\uq)\int^1_0\ud
y\int\frac{\ud^2\uQcal_\perp}{(2\pi)^2}\left[\frac{\uQcal^2_\perp+M^2(2y-1)^2}{(\uQcal^2_\perp+M^2+y(1-y)\uq^2)^2}-(M\to
M_\textrm{PV})\right],
\\
G_{3\sigma}(q_z,\uq_\perp)&=\frac{q_z}{|\uq|}\,\Pi(\uq)\Sigma(\uq)\int^1_0\ud
y\int\frac{\ud^2\uQcal_\perp}{(2\pi)^2}\left[\frac{\uQcal^2_\perp+M^2(2y-1)}{(\uQcal^2_\perp+M^2+y(1-y)\uq^2)^2}-(M\to
M_\textrm{PV})\right],
\end{align}
\end{subequations}
where $q_z=z\uM=(z_4+z_5)\uM$ and
$\uq_\perp=\up_{4\perp}+\up_{5\perp}$.

There are three integrals in the vector case
\begin{equation}
K^V_{\pi\pi},\,K^V_{33},\,K^V_{\sigma\sigma}
\end{equation} and four in
the axial one
\begin{equation}
K^A_{\pi\pi},\,K^A_{33},\,K^A_{\sigma\sigma},\,K^V_{3\sigma}.
\end{equation}
The contribution of the sea quark or antiquark to the axial charges
is obtained when the axial current probes the sea pair. This
contribution is proportional to $K^V_{3\sigma}$ which can be
understood as follows: the axial operator acting on a
quark-antiquark pair triggers a transition between the scalar
$\Sigma$ and pseudoscalar $\Pi$ pair configurations as denoted by
the subscript $3\sigma$ while the valence quarks remain unaffected
as denoted by the superscript $V$.\newline The contribution of
valence quarks to the axial charges is obtained when the axial
current probes the valence quark. This contribution is a linear
combination of $K^A_{\pi\pi}$, $K^A_{\sigma\sigma}$ and $K^A_{33}$
which can be understood as follows: the quark-antiquark pair stays
in a scalar or pseudoscalar configuration as denoted by the
subscripts $\pi\pi,33,\sigma\sigma$ but now the probe sees the axial
valence probability distribution $\Phi^A$ as denoted by the
superscript $A$.

\subsection{$7Q$ scalar integrals}

In the $7Q$ sector there are two quark-antiquark pairs and twenty
integrals appear after contractions. These integrals can be written
in the general form
\begin{equation}
K^I_J=\frac{M^4}{(2\pi)^2}\int \frac{\ud^3 \uq}{(2\pi)^3}\frac{\ud^3
\uq'}{(2\pi)^3}\,\Phi^I\left(\frac{(q_z+q'_z)}{\uM},\uq_\perp+\uq'_\perp\right)\theta(q_z)\,\theta(q'_z)\,q_z\,q'_z\,G_J(q_z,q'_z,\uq_\perp,\uq'_\perp),
\end{equation}
where $J=\pi\pi\pi\pi,\pi\pi\pi\pi 2,\pi\pi 33,3333,\pi 3\pi
3,\sigma\sigma\pi\pi,\sigma\sigma 33,\sigma\sigma\sigma\sigma,\pi\pi
3\sigma,333\sigma,\pi 3\pi\sigma,\sigma\sigma 3\sigma$. These
distributions are obtained by contracting four quark-antiquark wave
functions $W$, see eq. \eqref{Pair wavefunction} and regularized by
means of Pauli-Villars procedure. They can be expressed in terms of
$G_J(q_z,\uq_\perp)$. Here are then the distributions in the $7Q$
sector
\begin{subequations}\label{int}
\begin{align}
G_{\pi\pi\pi\pi}(q_z,q'_z,\uq_\perp,\uq'_\perp)&=G_{\pi\pi}(q_z,\uq_\perp)\,G_{\pi\pi}(q'_z,\uq'_\perp),
\\
G_{\pi\pi\pi\pi
2}(q_z,q'_z,\uq_\perp,\uq'_\perp)&=\frac{(\uq\cdot\uq')^2}{\uq^2\uq'^2}\,G_{\pi\pi}(q_z,\uq_\perp)\,G_{\pi\pi}(q'_z,\uq'_\perp),
\\
G_{\pi\pi
33}(q_z,q'_z,\uq_\perp,\uq'_\perp)&=G_{\pi\pi}(q_z,\uq_\perp)\,G_{33}(q'_z,\uq'_\perp),
\\
G_{3333}(q_z,q'_z,\uq_\perp,\uq'_\perp)&=G_{33}(q_z,\uq_\perp)\,G_{33}(q'_z,\uq'_\perp),
\\
G_{\pi 3\pi
3}(q_z,q'_z,\uq_\perp,\uq'_\perp)&=\frac{q_zq'_z(\uq\cdot\uq')}{\uq^2\uq'^2}\,G_{\pi\pi}(q_z,\uq_\perp)\,G_{\pi\pi}(q'_z,\uq'_\perp),
\\
G_{\sigma\sigma\pi\pi}(q_z,q'_z,\uq_\perp,\uq'_\perp)&=G_{\sigma\sigma}(q_z,\uq_\perp)\,G_{\pi\pi}(q'_z,\uq'_\perp),
\\
G_{\sigma\sigma
33}(q_z,q'_z,\uq_\perp,\uq'_\perp)&=G_{\sigma\sigma}(q_z,\uq_\perp)\,G_{33}(q'_z,\uq'_\perp),
\\
G_{\sigma\sigma\sigma\sigma}(q_z,q'_z,\uq_\perp,\uq'_\perp)&=G_{\sigma\sigma}(q_z,\uq_\perp)\,G_{\sigma\sigma}(q'_z,\uq'_\perp),
\\
G_{\pi\pi
3\sigma}(q_z,q'_z,\uq_\perp,\uq'_\perp)&=G_{\pi\pi}(q_z,\uq_\perp)\,G_{3\sigma}(q'_z,\uq'_\perp),
\\
G_{333\sigma}(q_z,q'_z,\uq_\perp,\uq'_\perp)&=G_{33}(q_z,\uq_\perp)\,G_{3\sigma}(q'_z,\uq'_\perp),
\\
G_{\pi
3\pi\sigma}(q_z,q'_z,\uq_\perp,\uq'_\perp)&=\frac{q_z(\uq\cdot\uq')}{q'_z\uq^2}\,G_{\pi\pi}(q_z,\uq_\perp)\,G_{3\sigma}(q'_z,\uq'_\perp),
\\
G_{\sigma\sigma
3\sigma}(q_z,q'_z,\uq_\perp,\uq'_\perp)&=G_{\sigma\sigma}(q_z,\uq_\perp)\,G_{3\sigma}(q'_z,\uq'_\perp),
\end{align}
\end{subequations}
where $q_z=z\uM=(z_4+z_5)\uM$, $q'_z=z\uM=(z_6+z_7)\uM$,
$\uq_\perp=\up_{4\perp}+\up_{5\perp}$ and
$\uq'_\perp=\up_{6\perp}+\up_{7\perp}$.\newline There are eight
integrals in the vector case
\begin{equation}
K^V_{\pi\pi\pi\pi},\,K^V_{\pi\pi\pi\pi 2},\,K^V_{\pi\pi
33},\,K^V_{3333},\,K^V_{\pi 3\pi
3},\,K^V_{\sigma\sigma\pi\pi},\,K^V_{\sigma\sigma 33},\,
K^V_{\sigma\sigma\sigma\sigma}
\end{equation}and twelve in the axial
one
\begin{equation}
K^A_{\pi\pi\pi\pi},\,K^A_{\pi\pi\pi\pi 2},\,K^A_{\pi\pi
33},\,K^A_{3333},\,K^A_{\pi 3\pi
3},\,K^A_{\sigma\sigma\pi\pi},\,K^A_{\sigma\sigma
33},\,K^A_{\sigma\sigma\sigma\sigma},\,K^V_{\pi\pi
3\sigma},\,K^V_{333\sigma},\,K^V_{\pi
3\pi\sigma},\,K^V_{\sigma\sigma 3\sigma}.
\end{equation}
The contribution of the sea quarks or antiquarks to the axial
charges is a linear combination of $K^V_{\pi\pi 3\sigma}$,
$K^V_{333\sigma}$, $K^V_{\pi 3\pi\sigma}$ and $K^V_{\sigma\sigma
3\sigma}$. There are more integrals than in the $5Q$ case since the
undisturbed quark-antiquark pair is either in a scalar
($\usigma\usigma 3\sigma$) or pseudoscalar combination ($\upi\upi
3\sigma,\mathbf{33}3\sigma,\upi\mathbf{3}\pi\sigma$).\newline The
contribution of valence quarks to the axial charges is a linear
combination of $K^A_{\pi\pi\pi\pi}$, $K^A_{\pi\pi\pi\pi 2}$,
$K^A_{\pi\pi 33}$, $K^A_{3333}$, $K^A_{\pi 3\pi 3}$,
$K^A_{\sigma\sigma\pi\pi}$, $K^A_{\sigma\sigma 33}$ and
$K^A_{\sigma\sigma\sigma\sigma}$. There are more integrals than in
the $5Q$ case since the undisturbed quark-antiquark pairs are in a
purely scalar ($\sigma\sigma\sigma\sigma$) or purely pseudoscalar
($\pi\pi\pi\pi,\pi\pi\pi\pi 2,\pi\pi 33,3333,\pi 3\pi 3$) or mixed
combination ($\sigma\sigma\pi\pi,\sigma\sigma 33$).

\section{Symmetry relations and parametrization}\label{Section cinq}

In this work we have studied vector and axial charges in flavor
$SU(3)$ symmetry. Even though this symmetry is broken in nature, it
gives quite a good estimation. Assuming a symmetry has the advantage
that all particles belonging to the same representation of the
symmetry are on the same footing and are related through pure
symmetry transformations. This means that for flavor $SU(3)$
symmetry, it is sufficient to consider only, say, the proton to
describe the whole baryon octet. Properties of the other members can
be obtained from those of the proton provided that flavor $SU(3)$
symmetry is considered.

The naive nonrelativistic quark model is based on a larger symmetry
group $SU(6)$ that imbeds $SU(3)\times SU(2)$. In this approach,
octet and decuplet baryons now belong to the same supermultiplet.
This yields relations \emph{between} different $SU(3)$ multiplets
and new ones within $SU(3)$ multiplets.

\subsection{$SU(3)$ relations for octet baryons}

As we have seen in the previous section, valence quark orbital
angular momentum just introduces a factor to charges and so $SU(6)$
symmetry is not broken. However, additional quark-antiquark pairs
break $SU(6)$ symmetry and therefore spoil NQM relations. Since the
present approach is based on flavor $SU(3)$ symmetry, we naturally
recover the expected relations imposed by this symmetry. In
principle, if we can determine the individual contributions of $u$,
$d$ and $s$ flavors to proton charges ($Q^u_p$, $Q^d_p$ and
$Q^s_p$), there is no need to go through the whole calculation once
more to determine the charges of other members of the octet. We
therefore expect that, for each charge, we need to know only three
quantities.

Experimentally, we do not have a direct access to the flavor
contributions of a given baryon charge. Nevertheless, under flavor
$SU(3)$ symmetry assumptions, one can extract from data combinations
of $Q^u_B$, $Q^d_B$ and $Q^s_B$, for a given baryon $B$
\begin{subequations}\label{chargecombinations}
\begin{align}
Q^{(3)}_B&=Q^u_B-Q^d_B&(\textrm{isovector})\rule{0pt}{3ex}\\
Q^{(8)}_B&=\left(Q^u_B+Q^d_B-2Q^s_B\right)/\sqrt{3}&(\textrm{octet})\rule{0pt}{3ex}\\
Q^{(0)}_B&=Q^u_B+Q^d_B+Q^s_B&(\textrm{singlet})\rule{0pt}{3ex}
\end{align}
\end{subequations}
Because of flavor $SU(3)$ symmetry, these charges can be expressed
as linear combinations of $Q^{u,d,s}_p$ for any octet baryon $B$.
Naturally, if one knows \emph{all} axial charges of a \emph{given}
baryon $B$, by inversion of \eqref{chargecombinations}, one can
extract $Q^{u,d,s}_B$. Except for $\Lambda_\mathbf{8}^0$ and
$\Sigma_\mathbf{8}^0$, one can also extract $Q^{u,d,s}_p$, even if
$B\neq p$. One could also try to extract $Q^{u,d,s}_p$ from a
\emph{given} axial charge, say $Q^{(3)}_B$, of \emph{all} octet
baryons. This is in fact not sufficient. To see this, we have used
the projection technique on quantum numbers shortly described in
subsection \ref{Projection}. It allowed us to write the contribution
of any flavor to a charge\footnote{This has been done for vector,
axial and tensor charges.} of any octet baryon in terms of linear
combinations of $K$ integrals. It was then possible to write
$Q^{u,d,s}_B$ for any octet baryon $B$ in terms of $Q^{u,d,s}_p$
which are our desired $SU(3)$ relations. Instead of using
$Q^{u,d,s}_p$ we prefer to use three other quantities $\alpha$,
$\beta$ and $\gamma$. The expressions for $Q^{u,d,s}_B$ in terms of
$\alpha$, $\beta$ and $\gamma$ can be found in Table
\ref{Octetcontent}.
\begin{table}[h!]\begin{center}\caption{\small{$SU(3)$
octet relations.\newline}}
\begin{tabular}{c|ccc} \hline\hline
$B$&$Q^u_B$&$Q^d_B$&$Q^s_B$\rule{0pt}{3ex}\\\hline \rule{0pt}{3ex}
$p_\mathbf{8}^+$&$\alpha+\gamma$&$\beta+\gamma$&$\gamma$\\\rule{0pt}{3ex}
$n_\mathbf{8}^0$&$\beta+\gamma$&$\alpha+\gamma$&$\gamma$\\\rule{0pt}{3ex}
$\Lambda_\mathbf{8}^0$&$\frac{1}{6}(\alpha+4\beta)+\gamma$&$\frac{1}{6}(\alpha+4\beta)+\gamma$&$\frac{1}{3}(2\alpha-\beta)+\gamma$\\\rule{0pt}{3ex}
$\Sigma_\mathbf{8}^+$&$\alpha+\gamma$&$\gamma$&$\beta+\gamma$\\\rule{0pt}{3ex}
$\Sigma_\mathbf{8}^0$&$\frac{1}{2}\,\alpha+\gamma$&$\frac{1}{2}\,\alpha+\gamma$&$\beta+\gamma$\\\rule{0pt}{3ex}
$\Sigma_\mathbf{8}^-$&$\gamma$&$\alpha+\gamma$&$\beta+\gamma$\\\rule{0pt}{3ex}
$\Xi_\mathbf{8}^0$&$\beta+\gamma$&$\gamma$&$\alpha+\gamma$\\\rule{0pt}{3ex}
$\Xi_\mathbf{8}^-$&$\gamma$&$\beta+\gamma$&$\alpha+\gamma$\rule[-2ex]{0pt}{5ex}\\\hline\hline
\end{tabular}\label{Octetcontent}\end{center}
\end{table}
Note that we have also observed that these expressions still hold
separately for valence quarks, sea quarks and antiquarks.

The reason why we choose the set $\{\alpha,\beta,\gamma\}$ is
motivated by the fact that it makes obvious the statement that we
cannot extract univocally the flavor contributions of any octet
baryon by means of charges $Q^{(i)}_B$ ($i=3,8$ or 0) for all octet
baryons $B$. Indeed, the isovector (3) and octet (8) combinations do
not depend on $\gamma$ as one can directly see from the definitions
(\ref{chargecombinations}) and Table \ref{Octetcontent}. Concerning
the isosinglet combination (0), one can directly notice that it has
the same value for all members of the octet
\begin{equation}
Q^{(0)}_B=\alpha+\beta+3\gamma.
\end{equation}

A few octet baryon decay constants are known experimentally. It is
then useful to express them in terms of our parameters $\alpha$ and
$\beta$ ($\gamma$ disappears as explained earlier), see Table
\ref{axialdecay}.
\begin{table}[h!]\begin{center}\caption{\small{$SU(3)$
octet transition relations.\newline}}
\begin{tabular}{r@{$\to$\,}l|c||r@{$\to$\,}l|c} \hline\hline
\multicolumn{2}{c|}{Transitions}&$g_{V,A}$&\multicolumn{2}{c|}{Transitions}&$g_{V,A}$\rule{0pt}{3ex}\\\hline
\rule{0pt}{3ex}
$n_\mathbf{8}^0$&$p_\mathbf{8}^+$&$\alpha-\beta$&$\textrm{
}\Sigma_\mathbf{8}^-$&$n_\mathbf{8}^0$&$-\beta$\\\rule{0pt}{3ex}
$\Sigma_\mathbf{8}^-$&$\Sigma_\mathbf{8}^0$&$\alpha/\sqrt{2}$&$\Xi_\mathbf{8}^-$&$\Sigma_\mathbf{8}^0$&$(\beta-\alpha)/\sqrt{2}$\\\rule{0pt}{3ex}
$\Sigma_\mathbf{8}^-$&$\Lambda_\mathbf{8}^0$&$(\alpha-2\beta)/\sqrt{6}$&$\Xi_\mathbf{8}^-$&$\Lambda_\mathbf{8}^0$&$-(\alpha+\beta)/\sqrt{6}$\\\rule{0pt}{3ex}
$\Sigma_\mathbf{8}^0$&$\Sigma_\mathbf{8}^+$&$-\alpha/\sqrt{2}$&$\Sigma_\mathbf{8}^0$&$p_\mathbf{8}^+$&$-\beta/\sqrt{2}$\\\rule{0pt}{3ex}
$\Lambda_\mathbf{8}^0$&$\Sigma_\mathbf{8}^+$&$(\alpha-2\beta)/\sqrt{6}$&$\Lambda_\mathbf{8}^0$&$p_\mathbf{8}^+$&$(\beta-2\alpha)/\sqrt{6}$\\\rule{0pt}{3ex}
$\Xi_\mathbf{8}^-$&$\Xi_\mathbf{8}^0$&$\beta$&$\Xi_\mathbf{8}^0$&$\Sigma_\mathbf{8}^+$&$\alpha-\beta$\rule[-2ex]{0pt}{5ex}\\\hline\hline
\end{tabular}\label{axialdecay}\end{center}
\end{table}
In the literature, one often uses another set of two parameters to
describe all these octet transitions, known as the $F\& D$ Cabibbo
parameters \cite{Cabibbo}. These parameters can be related to our
$\alpha$ and $\beta$ by means of the relations
\begin{equation}
\alpha=2F,\qquad \beta=F-D.
\end{equation}

It is also interesting to consider the limit where baryons are made
of 3 quarks only ($3Q$). Using the projection technique, this
corresponds to taking $\gamma=0$. In this case, protons are only
made of $u$ and $d$ quarks as expected and there are only valence
quarks. At the $5Q$ level, $\gamma\neq 0$ and we obtain Table
\ref{Octetcontent}. The $7Q$ component does not change anything
concerning the $SU(3)$ relations and we may reasonably expect that
it would also be the case for any additional quark-antiquark pair.
As a last remark concerning octet baryons, we would like to stress
that we naturally obtain that the strange contribution in the proton
is the same as the strange contribution in the neutron
\begin{equation}
Q^s_p=Q^s_n=\gamma.
\end{equation}
In fact, all members of a given isomultiplet ($N,\Lambda,\Sigma$ or
$\Xi$) have the same strange contribution to the charges.

\subsection{$SU(3)$ relations for decuplet baryons}

The same game has been done for decuplet baryons. For this
multiplet, we in fact observed that only two parameters, say
$\alpha'$ and $\beta'$, are necessary. The expressions for
$Q^{u,d,s}_B$ in terms of $\alpha'$ and $\beta'$ can be found in
Table \ref{Decupletcontent}.
\begin{table}[h!]\begin{center}\caption{\small{$SU(3)$ decuplet relations.\newline}}
\begin{tabular}{c|ccc}
\hline\hline $B$&$Q^u_B$&$Q^d_B$&$Q^s_B$\rule{0pt}{3ex}\\\hline
\rule{0pt}{3ex}
$\Delta_\mathbf{10}^{++}$&$3\alpha'+\beta'$&$\beta'$&$\beta'$\\\rule{0pt}{3ex}
$\Delta_\mathbf{10}^+$&$2\alpha'+\beta'$&$\alpha'+\beta'$&$\beta'$\\\rule{0pt}{3ex}
$\Delta_\mathbf{10}^0$&$\alpha'+\beta'$&$2\alpha'+\beta'$&$\beta'$\\\rule{0pt}{3ex}
$\Delta_\mathbf{10}^-$&$\beta'$&$3\alpha'+\beta'$&$\beta'$\\\rule{0pt}{3ex}
$\Sigma_\mathbf{10}^+$&$2\alpha'+\beta'$&$\beta'$&$\alpha'+\beta'$\\\rule{0pt}{3ex}
$\Sigma_\mathbf{10}^0$&$\alpha'+\beta'$&$\alpha'+\beta'$&$\alpha'+\beta'$\\\rule{0pt}{3ex}
$\Sigma_\mathbf{10}^-$&$\beta'$&$2\alpha'+\beta'$&$\alpha'+\beta'$\\\rule{0pt}{3ex}
$\Xi_\mathbf{10}^0$&$\alpha'+\beta'$&$\beta'$&$2\alpha'+\beta'$\\\rule{0pt}{3ex}
$\Xi_\mathbf{10}^-$&$\beta'$&$\alpha'+\beta'$&$2\alpha'+\beta'$\\\rule{0pt}{3ex}
$\Omega_\mathbf{10}^-$&$\beta'$&$\beta'$&$3\alpha'+\beta'$\rule[-2ex]{0pt}{5ex}\\
\hline\hline
\end{tabular}\label{Decupletcontent}\end{center}
\end{table}

Once more, one cannot extract the flavor contribution of one
decuplet baryon from the knowledge of a charge $Q^{(i)}_B$ with
$i=3,8$ or 0 for all decuplet baryons. In the $3Q$ limit, we have
$\beta'=0$ while in presence of quark-antiquark pairs $\beta'\neq
0$. Moreover, all the members of a given isomultiplet
($\Delta,\Sigma,\Xi$ or $\Omega$) have the same strange contribution
to the charges. Finally, all the members of the decuplet have the
same isosinglet contribution
\begin{equation}
Q^{(0)}_B=3\left(\alpha'+\beta'\right).
\end{equation}
\newpage
\subsection{$SU(3)$ relations for antidecuplet baryons}

The antidecuplet is very similar to the decuplet. Here also only two
parameters are sufficient, say $\alpha''$ and $\beta''$, and yield
the relations in Table \ref{Antidecupletcontent}.
\begin{table}[h!]\begin{center}\caption{\small{$SU(3)$ antidecuplet relations.\newline}}
\begin{tabular}{c|ccc}
\hline\hline $B$&$Q^u_B$&$Q^d_B$&$Q^s_B$\rule{0pt}{3ex}\\\hline
\rule{0pt}{3ex}
$\Theta_\mathbf{\overline{10}}^+$&$2\alpha''+\beta''$&$2\alpha''+\beta''$&$-\alpha''+\beta''$\\\rule{0pt}{3ex}
$p_\mathbf{\overline{10}}^+$&$2\alpha''+\beta''$&$\alpha''+\beta''$&$\beta''$\\\rule{0pt}{3ex}
$n_\mathbf{\overline{10}}^0$&$\alpha''+\beta''$&$2\alpha''+\beta''$&$\beta''$\\\rule{0pt}{3ex}
$\Sigma_\mathbf{\overline{10}}^+$&$2\alpha''+\beta''$&$\beta''$&$\alpha''+\beta''$\\\rule{0pt}{3ex}
$\Sigma_\mathbf{\overline{10}}^0$&$\alpha''+\beta''$&$\alpha''+\beta''$&$\alpha''+\beta''$\\\rule{0pt}{3ex}
$\Sigma_\mathbf{\overline{10}}^-$&$\beta''$&$2\alpha''+\beta''$&$\alpha''+\beta''$\\\rule{0pt}{3ex}
$\Xi_\mathbf{\overline{10}}^+$&$2\alpha''+\beta''$&$-\alpha''+\beta''$&$2\alpha''+\beta''$\\\rule{0pt}{3ex}
$\Xi_\mathbf{\overline{10}}^0$&$\alpha''+\beta''$&$\beta''$&$2\alpha''+\beta''$\\\rule{0pt}{3ex}
$\Xi_\mathbf{\overline{10}}^-$&$\beta''$&$\alpha''+\beta''$&$2\alpha''+\beta''$\\\rule{0pt}{3ex}
$\Xi_\mathbf{\overline{10}}^{--}$&$-\alpha''+\beta''$&$2\alpha''+\beta''$&$2\alpha''+\beta''$\rule[-2ex]{0pt}{5ex}\\
\hline\hline
\end{tabular}\label{Antidecupletcontent}\end{center}
\end{table}

The relations are different but the comments made for the decuplet
also apply to the antidecuplet (excepted that the $'$ are replaced
by $''$) excepted that the $3Q$ limit does not exist since
pentaquarks involve at least one quark-antiquark pair. The $3Q$
contribution is indeed identically zero when using the projection
technique.

\section{Results}\label{Section six}

In this section we present our results. In the following we give the
expressions for the octet, decuplet and antidecuplet normalizations,
vector and axial parameters
$\alpha,\beta,\gamma,\alpha',\beta',\alpha'',\beta''$ in terms of
the scalar overlap integrals $K$ in the $3Q$, $5Q$ and $7Q$ sectors.
Note that we do not give the $7Q$ sector for decuplet baryons. While
we have all ingredients, the contractions involved are too complex
and too long to be computed in a reasonable amount of time. We
finally give the numerical evaluation of the scalar overlap
integrals and collect in tables all our pre- and postdictions.

We split the contribution to the charges into valence quark, sea
quark and antiquark contributions, \emph{i.e.} we have in the vector
case
\begin{equation}
q_\textrm{tot}=q_\textrm{val}+q_\textrm{sea}-\bar q
\end{equation}
and in the axial one
\begin{equation}
\Delta q_\textrm{tot}=\Delta q_\textrm{val}+\Delta
q_\textrm{sea}+\Delta\bar q,
\end{equation}
where ``val'' refers to the valence quarks and ``sea'' to the sea
quarks.
\newline The vector charges can be understood as follows:
they count the total number of quarks
($q_\textrm{val,sea}=q_{\textrm{val,sea}+}+q_{\textrm{val,sea}-}$)
\emph{minus} the total number of antiquarks ($\bar q=\bar q_++\bar
q_-$), irrespective of their polarization. The vector charges $\bar
q\gamma^+ q$ then just give the effective number of quarks of flavor
$q=u,d,s$ in the baryon.\newline The axial charges count the total
number of constituents with polarization parallel \emph{minus} the
total number of constituents with polarization antiparallel to the
baryon longitudinal polarization, irrespective of their quark
($\Delta
q_{\textrm{val,sea}}=q_{\textrm{val,sea}+}-q_{\textrm{val,sea}-}$)
or antiquark nature ($\Delta\bar q=\bar q_+-\bar q_-$). The axial
charges $\bar q\gamma^+\gamma^5 q$ then give the contribution of
quarks of flavor $q=u,d,s$ to the total baryon longitudinal
polarization.

We would like to stress here a somewhat confusing point. In this
paper, we call ``valence'' quarks those populating the discrete
level of the spectrum \eqref{Discrete level IMF}. Our valence
contribution to charges is in fact the discrete level contribution.
In the literature, the valence contribution refers to the
\emph{effective} contribution $q_v$ or $\Delta q_v$, which
corresponds to the contribution of all quarks \emph{minus} the
contribution of all antiquarks. In this sense, this is the reason
why one often says that only valence quarks contribute to vector
charges $q_\textrm{tot}=q_v$ while both valence quarks and
quark-antiquark pairs contribute to the axial ones $\Delta
q_\textrm{tot}=\Delta q_v+2\Delta\bar q$. This point of view is
based on the perturbative picture of the nucleon sea. Indeed, in
this picture, quark-antiquark pairs are generated by gluon splitting
leading to the equality of the sea quark and antiquark
contributions. Only in this picture can our meaning of valence
quarks and the literature one be identified. In a non-perturbative
picture, the sea quark and antiquark contributions are not forced to
be equal anymore. Consequently we have in general
$q_\textrm{val}\neq q_v$ and $\Delta q_\textrm{val}\neq\Delta q_v$.

\subsection{Octet baryons}

Here are the expressions for the octet baryons. They are obtained by
contracting the octet baryon wave functions without any charge
acting on the quark lines. The upper indices $3,5,7$ refer to the
$3Q$, $5Q$ and $7Q$ Fock sectors.\newline The contributions to the
octet normalization are
\begin{subequations}
\begin{align}
\uN^{(3)}(B_{\bf 8})&=9\,\Phi^V(0,0),\\
\uN^{(5)}(B_{\bf
8})&=\frac{18}{5}\left(11K^V_{\pi\pi}+23K^V_{\sigma\sigma}\right),\\
\uN^{(7)}(B_{\bf
8})&=\frac{144}{5}\left(15K^V_{\pi\pi\pi\pi}+5K^V_{\pi\pi\pi\pi
2}+52K^V_{\sigma\sigma\pi\pi}+54K^V_{\sigma\sigma\sigma\sigma}\right).
\end{align}
\end{subequations}
In the $3Q$ sector there is no quark-antiquark pair and thus only
valence quarks contribute to the charges
\begin{align}
\alpha_{V,q_\textrm{val}}^{(3)}&=18\,\Phi^V(0,0),
&\beta_{V,q_\textrm{val}}^{(3)}&=9\,\Phi^V(0,0),
&\gamma_{V,q_\textrm{val}}^{(3)}&=0,\\
\alpha_{A,q_\textrm{val}}^{(3)}&=12\,\Phi^A(0,0),
&\beta_{A,q_\textrm{val}}^{(3)}&=-3\,\Phi^A(0,0),
&\gamma_{A,q_\textrm{val}}^{(3)}&=0.
\end{align}
In the $5Q$ sector one has
\begin{subequations}
\begin{align}
\alpha_{V,q_\textrm{val}}^{(5)}&=\frac{18}{5}\left(15K^V_{\pi\pi}+43K^V_{\sigma\sigma}\right),&
\alpha_{V,q_\textrm{sea}}^{(5)}&=\frac{132}{5}\left(K^V_{\pi\pi}+K^V_{\sigma\sigma}\right),&
\alpha_{V,\bar
q}^{(5)}&=\frac{6}{5}\left(K^V_{\pi\pi}+13K^V_{\sigma\sigma}\right),\\
\beta_{V,q_\textrm{val}}^{(5)}&=\frac{72}{25}\left(12K^V_{\pi\pi}+25K^V_{\sigma\sigma}\right),&
\beta_{V,q_\textrm{sea}}^{(5)}&=\frac{24}{25}\left(13K^V_{\pi\pi}+22K^V_{\sigma\sigma}\right),&
\beta_{V,\bar
q}^{(5)}&=\frac{6}{25}\left(31K^V_{\pi\pi}+43K^V_{\sigma\sigma}\right),\\
\gamma_{V,q_\textrm{val}}^{(5)}&=\frac{36}{25}\left(7K^V_{\pi\pi}+5K^V_{\sigma\sigma}\right),&
\gamma_{V,q_\textrm{sea}}^{(5)}&=\frac{6}{25}\left(K^V_{\pi\pi}+49K^V_{\sigma\sigma}\right),&
\gamma_{V,\bar
q}^{(5)}&=\frac{6}{25}\left(43K^V_{\pi\pi}+79K^V_{\sigma\sigma}\right),
\end{align}
\end{subequations}
\begin{subequations}
\begin{align}
\alpha_{A,q_\textrm{val}}^{(5)}&=\frac{6}{5}\left(29K^A_{\pi\pi}+2K^A_{33}+91K^A_{\sigma\sigma}\right),&
\alpha_{A,q_\textrm{sea}}^{(5)}&=\frac{-168}{5}\,K^A_{3\sigma},&
\alpha_{A,\bar q}^{(5)}&=\frac{-132}{5}\,K^A_{3\sigma},\\
\beta_{A,q_\textrm{val}}^{(5)}&=\frac{-24}{25}\left(16K^A_{\pi\pi}-11K^A_{33}+26K^A_{\sigma\sigma}\right),&
\beta_{A,q_\textrm{sea}}^{(5)}&=\frac{408}{25}\,K^A_{3\sigma},&
\beta_{A,\bar q}^{(5)}&=\frac{228}{25}\,K^A_{3\sigma},\\
\gamma_{A,q_\textrm{val}}^{(5)}&=\frac{-12}{25}\left(11K^A_{\pi\pi}-16K^A_{33}+K^A_{\sigma\sigma}\right),&
\gamma_{A,q_\textrm{sea}}^{(5)}&=\frac{84}{25}\,K^A_{3\sigma},&
\gamma_{A,\bar
q}^{(5)}&=\frac{84}{25}\,K^A_{3\sigma}.\label{octet5Q}
\end{align}
\end{subequations}
In the $7Q$ sector one has
\begin{subequations}
\begin{align}
\alpha_{V,q_\textrm{val}}^{(7)}&=\frac{48}{5}\left(49K^V_{\pi\pi\pi\pi}+38K^V_{\pi\pi\pi\pi
2}+200K^V_{\sigma\sigma\pi\pi}+285K^V_{\sigma\sigma\sigma\sigma}\right),\\
\alpha_{V,q_\textrm{sea}}^{(7)}&=\frac{48}{5}\left(47K^V_{\pi\pi\pi\pi}+2K^V_{\pi\pi\pi\pi
2}+144K^V_{\sigma\sigma\pi\pi}+99K^V_{\sigma\sigma\sigma\sigma}\right),\\
\alpha_{V,\bar
q}^{(7)}&=\frac{96}{5}\left(3K^V_{\pi\pi\pi\pi}+5K^V_{\pi\pi\pi\pi
2}+16K^V_{\sigma\sigma\pi\pi}+30K^V_{\sigma\sigma\sigma\sigma}\right),\\
\beta_{V,q_\textrm{val}}^{(7)}&=\frac{48}{25}\left(181K^V_{\pi\pi\pi\pi}+41K^V_{\pi\pi\pi\pi
2}+626K^V_{\sigma\sigma\pi\pi}+618K^V_{\sigma\sigma\sigma\sigma}\right),\\
\beta_{V,q_\textrm{sea}}^{(7)}&=\frac{96}{25}\left(61K^V_{\pi\pi\pi\pi}+22K^V_{\pi\pi\pi\pi
2}+201K^V_{\sigma\sigma\pi\pi}+198K^V_{\sigma\sigma\sigma\sigma}\right),\\
\beta_{V,\bar
q}^{(7)}&=\frac{96}{25}\left(39K^V_{\pi\pi\pi\pi}+5K^V_{\pi\pi\pi\pi
2}+124K^V_{\sigma\sigma\pi\pi}+102K^V_{\sigma\sigma\sigma\sigma}\right),\\
\gamma_{V,q_\textrm{val}}^{(7)}&=\frac{48}{25}\left(83K^V_{\pi\pi\pi\pi}-2K^V_{\pi\pi\pi\pi
2}+238K^V_{\sigma\sigma\pi\pi}+129K^V_{\sigma\sigma\sigma\sigma}\right),\\
\gamma_{V,q_\textrm{sea}}^{(7)}&=\frac{48}{25}\left(31K^V_{\pi\pi\pi\pi}+32K^V_{\pi\pi\pi\pi
2}+146K^V_{\sigma\sigma\pi\pi}+243K^V_{\sigma\sigma\sigma\sigma}\right),\\
\gamma_{V,\bar
q}^{(7)}&=\frac{288}{25}\left(19K^V_{\pi\pi\pi\pi}+5K^V_{\pi\pi\pi\pi
2}+64K^V_{\sigma\sigma\pi\pi}+62K^V_{\sigma\sigma\sigma\sigma}\right),
\end{align}
\end{subequations}
\begin{subequations}
\begin{align}
\alpha_{A,q_\textrm{val}}^{(7)}&=\frac{48}{5}\left(33K^A_{\pi\pi\pi\pi}+30K^A_{\pi\pi\pi\pi
2}-2K^A_{\pi\pi 33}+4K^A_{\pi 3\pi 3}+134K^A_{\sigma\sigma\pi\pi}+10K^A_{\sigma\sigma 33}+211K^A_{\sigma\sigma\sigma\sigma}\right),\\
\alpha_{A,q_\textrm{sea}}^{(7)}&=\frac{-96}{5}\left(32K^A_{\pi\pi
3\sigma}-K^A_{\pi
3\pi\sigma}+65K^A_{\sigma\sigma 3\sigma}\right),\\
\alpha_{A,\bar q}^{(7)}&=\frac{-96}{5}\left(25K^A_{\pi\pi
3\sigma}+K^A_{\pi
3\pi\sigma}+52K^A_{\sigma\sigma 3\sigma}\right),\\
\beta_{A,q_\textrm{val}}^{(7)}&=\frac{-48}{25}\left(51K^A_{\pi\pi\pi\pi}+45K^A_{\pi\pi\pi\pi
2}+38K^A_{\pi\pi 33}-82K^A_{\pi 3\pi 3}+292K^A_{\sigma\sigma\pi\pi}-214K^A_{\sigma\sigma 33}+224K^A_{\sigma\sigma\sigma\sigma}\right),\\
\beta_{A,q_\textrm{sea}}^{(7)}&=\frac{192}{25}\left(35K^A_{\pi\pi
3\sigma}+2K^A_{\pi
3\pi\sigma}+77K^A_{\sigma\sigma 3\sigma}\right),\\
\beta_{A,\bar q}^{(7)}&=\frac{96}{25}\left(47K^A_{\pi\pi
3\sigma}-K^A_{\pi
3\pi\sigma}+92K^A_{\sigma\sigma 3\sigma}\right),\\
\gamma_{A,q_\textrm{val}}^{(7)}&=\frac{-48}{25}\left(13K^A_{\pi\pi\pi\pi}+10K^A_{\pi\pi\pi\pi
2}+24K^A_{\pi\pi 33}-56K^A_{\pi 3\pi 3}+106K^A_{\sigma\sigma\pi\pi}-152K^A_{\sigma\sigma 33}+7K^A_{\sigma\sigma\sigma\sigma}\right),\\
\gamma_{A,q_\textrm{sea}}^{(7)}&=\frac{96}{25}\left(25K^A_{\pi\pi
3\sigma}-8K^A_{\pi
3\pi\sigma}+37K^A_{\sigma\sigma 3\sigma}\right),\label{Octet7Q}\\
\gamma_{A,\bar q}^{(7)}&=\frac{288}{25}\left(7K^A_{\pi\pi
3\sigma}-K^A_{\pi 3\pi\sigma}+12K^A_{\sigma\sigma
3\sigma}\right).\label{Octet7Q2}
\end{align}
\end{subequations}

One can easily check that the obvious sum rules for the proton
\begin{subequations}
\begin{align}
\int\ud x\,[u(x)-\bar u(x)]&=2,\\\int\ud x\,[d(x)-\bar d(x)]&=1,\\
\int\ud x\,[s(x)-\bar s(x)]&=0
\end{align}
\end{subequations}
are satisfied separately in each sector. They are translated in our
parametrization as follows
\begin{subequations}
\begin{align}
\alpha_{V,q_\textrm{val}}^{(i)}+\alpha_{V,q_\textrm{sea}}^{(i)}-\alpha_{V,\bar
q}^{(i)}&=2\uN^{(i)}(B_{\bf 8}),\\
\beta_{V,q_\textrm{val}}^{(i)}+\beta_{V,q_\textrm{sea}}^{(i)}-\beta_{V,\bar
q}^{(i)}&=\uN^{(i)}(B_{\bf 8}),\\
\gamma_{V,q_\textrm{val}}^{(i)}+\gamma_{V,q_\textrm{sea}}^{(i)}-\gamma_{V,\bar
q}^{(i)}&=0,
\end{align}
\end{subequations}
for any $i=3Q,5Q,7Q,\cdots$.

\subsection{Decuplet baryons}\label{shape}

Here are the expressions for the decuplet baryons. They are obtained
by contracting the decuplet baryon wave functions without any charge
acting on the quark lines. The upper indices $i=3,5$ refer to the
$3Q$ and $5Q$ Fock sectors while the lower ones $3/2,1/2$ refer to
the $z$-component of the decuplet baryon spin.\newline The
contributions to the decuplet normalization are
\begin{subequations}
\begin{align}
\uN^{(3)}_{3/2}(B_{\bf 10})&=\uN^{(3)}_{1/2}(B_{\bf 10})=\frac{18}{5}\,\Phi^V(0,0),\\
\uN^{(5)}_{3/2}(B_{\bf
10})&=\frac{9}{5}\left(15K^V_{\pi\pi}-6K^V_{33}+17K^V_{\sigma\sigma}\right),\\
\uN^{(5)}_{1/2}(B_{\bf
10})&=\frac{9}{5}\left(11K^V_{\pi\pi}+6K^V_{33}+17K^V_{\sigma\sigma}\right).
\end{align}
\end{subequations}
In the $3Q$ sector there is no quark-antiquark pair and thus only
valence quarks contribute to the charges
\begin{align}
\alpha'^{(3)}_{V,q_\textrm{val},3/2}&=\alpha'^{(3)}_{V,q_\textrm{val},1/2}=\frac{18}{5}\,\Phi^V(0,0),
&\beta'^{(3)}_{V,q_\textrm{val},3/2}&=\beta'^{(3)}_{V,q_\textrm{val},1/2}=0,\\
\alpha'^{(3)}_{A,q_\textrm{val},3/2}&=3\alpha'^{(3)}_{A,q_\textrm{val},1/2}=\frac{18}{5}\,\Phi^A(0,0),
&\beta'^{(3)}_{A,q_\textrm{val},3/2}&=3\beta'^{(3)}_{A,q_\textrm{val},1/2}=0.
\end{align}
In the $5Q$ sector one has
\begin{subequations}
\begin{align}
\alpha'^{(5)}_{V,q_\textrm{val},3/2}&=\frac{9}{20}\left(33K^V_{\pi\pi}-6K^V_{33}+67K^V_{\sigma\sigma}\right),&
\alpha'^{(5)}_{V,q_\textrm{val},1/2}&=\frac{9}{20}\left(29K^V_{\pi\pi}+6K^V_{33}+67K^V_{\sigma\sigma}\right),\\
\alpha'^{(5)}_{V,q_\textrm{sea},3/2}&=\frac{3}{20}\left(57K^V_{\pi\pi}-30K^V_{33}+19K^V_{\sigma\sigma}\right),&
\alpha'^{(5)}_{V,q_\textrm{sea},1/2}&=\frac{3}{20}\left(37K^V_{\pi\pi}+30K^V_{33}+19K^V_{\sigma\sigma}\right),\\
\alpha'^{(5)}_{V,\bar
q,3/2}&=\frac{-6}{5}\left(3K^V_{\pi\pi}-3K^V_{33}-2K^V_{\sigma\sigma}\right),&
\alpha'^{(5)}_{V,\bar
q,1/2}&=\frac{-6}{5}\left(K^V_{\pi\pi}+3K^V_{33}-2K^V_{\sigma\sigma}\right),\\
\beta'^{(5)}_{V,q_\textrm{val},3/2}&=\frac{9}{20}\left(27K^V_{\pi\pi}-18K^V_{33}+K^V_{\sigma\sigma}\right),&
\beta'^{(5)}_{V,q_\textrm{val},1/2}&=\frac{9}{20}\left(15K^V_{\pi\pi}+18K^V_{33}+K^V_{\sigma\sigma}\right),\\
\beta'^{(5)}_{V,q_\textrm{sea},3/2}&=\frac{3}{20}\left(3K^V_{\pi\pi}+6K^V_{33}+49K^V_{\sigma\sigma}\right),&
\beta'^{(5)}_{V,q_\textrm{sea},1/2}&=\frac{3}{20}\left(7K^V_{\pi\pi}-6K^V_{33}+49K^V_{\sigma\sigma}\right),\\
\beta'^{(5)}_{V,\bar
q,3/2}&=\frac{3}{5}\left(21K^V_{\pi\pi}-12K^V_{33}+13K^V_{\sigma\sigma}\right),&
\beta'^{(5)}_{V,\bar
q,1/2}&=\frac{3}{5}\left(13K^V_{\pi\pi}+12K^V_{33}+13K^V_{\sigma\sigma}\right),
\end{align}
\end{subequations}
\begin{subequations}
\begin{align}
\alpha'^{(5)}_{A,q_\textrm{val},3/2}&=\frac{9}{20}\left(43K^A_{\pi\pi}-16K^A_{33}+67K^A_{\sigma\sigma}\right),&
\alpha'^{(5)}_{A,q_\textrm{val},1/2}&=\frac{3}{20}\left(23K^A_{\pi\pi}+44K^A_{33}+67K^A_{\sigma\sigma}\right),\\
\alpha'^{(5)}_{A,q_\textrm{sea},3/2}&=\frac{-99}{100}\,K^A_{3\sigma},&
\alpha'^{(5)}_{A,q_\textrm{sea},1/2}&=\frac{-33}{100}\,K^A_{3\sigma},\\
\alpha'^{(5)}_{A,\bar q,3/2}&=\frac{-36}{5}\,K^A_{3\sigma},&
\alpha'^{(5)}_{A,\bar q,1/2}&=\frac{-12}{5}\,K^A_{3\sigma},\\
\beta'^{(5)}_{A,q_\textrm{val},3/2}&=\frac{-9}{20}\left(23K^A_{\pi\pi}-32K^A_{33}-K^A_{\sigma\sigma}\right),&
\beta'^{(5)}_{A,q_\textrm{val},1/2}&=\frac{-3}{20}\left(19K^A_{\pi\pi}-20K^A_{33}-K^A_{\sigma\sigma}\right),\\
\beta'^{(5)}_{A,q_\textrm{sea},3/2}&=\frac{63}{10}\,K^A_{3\sigma},&
\beta'^{(5)}_{A,q_\textrm{sea},1/2}&=\frac{21}{10}\,K^A_{3\sigma},&\\
\beta'^{(5)}_{A,\bar q,3/2}&=\frac{18}{5}\,K^A_{3\sigma},&
\beta'^{(5)}_{A,\bar q,1/2}&=\frac{6}{5}\,K^A_{3\sigma}.
\end{align}
\end{subequations}
The $7Q$ sector of the decuplet has not been computed due to its far
greater complexity.

One can easily check that the obvious sum rules for $\Delta^{++}$
\begin{subequations}
\begin{align}
\int\ud x\,[u(x)-\bar u(x)]&=3,\\
\int\ud x\,[d(x)-\bar d(x)]&=0,\\
\qquad \int\ud x\,[s(x)-\bar s(x)]&=0
\end{align}
\end{subequations}
are satisfied separately in each sector. They are translated in our
parametrization as follows
\begin{subequations}
\begin{align}
\alpha'^{(i)}_{V,q_\textrm{val},J}+\alpha'^{(i)}_{V,q_\textrm{sea},J}-\alpha'^{(i)}_{V,\bar
q,J}&=\uN^{(i)}_J(B_{\bf 10}),\\
\beta'^{(i)}_{V,q_\textrm{val},J}+\beta'^{(i)}_{V,q_\textrm{sea},J}-\beta'^{(i)}_{V,\bar
q,J}&=0,
\end{align}
\end{subequations}
for any $i=3Q,5Q,7Q,\cdots$ and $J=3/2,1/2$.

Let us emphasize an interesting observation. If the decuplet was
made of three quarks only, then one would have the following
relations between spin-3/2 and 1/2 contributions
\begin{equation}\label{sphericity}
V_{3/2}=V_{1/2}, \qquad A_{3/2}=3A_{1/2},
\end{equation}
where $V$ stands for any vector contribution and $A$ for any axial
one. This picture presents the $\Delta$ as a spherical particle.
Things change in the $5Q$ sector. One notices directly that the
relations are broken by a unique structure
$(3K^V_{33}-K^V_{\pi\pi})$ in the vector case and
$(3K^A_{33}-K^A_{\pi\pi})$ in the axial one. Going back to the
definition of those integrals this is in fact a structure like
$\int\ud^3q\,f(\uq)\,(3q_z^2-q^2)$. This naturally reminds the
expression of a quadrupole
\begin{equation}
Q_{ij}=\int\ud^3r\,\rho(\ur)\,(3r_ir_j-r^2\delta_{ij})
\end{equation}
specified to the component $i=j=z$. Remarkably the present approach
shows \emph{explicitly} that the pion field is responsible for the
deviation of the $\Delta$ from spherical symmetry.

\subsection{Antidecuplet baryons}

Here are the expressions for the antidecuplet baryons. They are
obtained by contracting the antidecuplet baryon wave functions
without any charge acting on the quark lines. The upper indices
$5,7$ refer to the $5Q$ and $7Q$ Fock sectors\footnote{We remind
that there is no $3Q$ component in pentaquarks.}.\newline The
contributions to the antidecuplet normalization are
\begin{subequations}
\begin{align}
\uN^{(5)}(B_{\bf\overline{10}})&=\frac{36}{5}\left(K^V_{\pi\pi}+K^V_{\sigma\sigma}\right),\\
\uN^{(7)}(B_{\bf\overline{10}})&=\frac{72}{5}\left(9K^V_{\pi\pi\pi\pi}+K^V_{\pi\pi\pi\pi
2}+26K^V_{\sigma\sigma\pi\pi}+18K^V_{\sigma\sigma\sigma\sigma}\right).
\end{align}
\end{subequations}
In the $5Q$ sector one has
\begin{subequations}
\begin{align}
\alpha''^{(5)}_{V,q_\textrm{val}}&=\frac{18}{5}\left(K^V_{\pi\pi}+K^V_{\sigma\sigma}\right),&
\alpha''^{(5)}_{V,q_\textrm{sea}}&=\frac{6}{5}\left(K^V_{\pi\pi}+K^V_{\sigma\sigma}\right),&
\alpha''^{(5)}_{V,\bar
q}&=\frac{-12}{5}\left(K^V_{\pi\pi}+K^V_{\sigma\sigma}\right),\\
\beta''^{(5)}_{V,q_\textrm{val}}&=\frac{18}{5}\left(K^V_{\pi\pi}+K^V_{\sigma\sigma}\right),&
\beta''^{(5)}_{V,q_\textrm{sea}}&=\frac{6}{5}\left(K^V_{\pi\pi}+K^V_{\sigma\sigma}\right),&
\beta''^{(5)}_{V,\bar
q}&=\frac{24}{5}\left(K^V_{\pi\pi}+K^V_{\sigma\sigma}\right),
\end{align}
\end{subequations}
\begin{subequations}
\begin{align}
\alpha''^{(5)}_{A,q_\textrm{val}}&=\frac{-6}{5}\left(K^A_{\pi\pi}-2K^A_{33}-K^A_{\sigma\sigma}\right),&
\alpha''^{(5)}_{A,q_\textrm{sea}}&=\frac{12}{5}\,K^A_{3\sigma},&
\alpha''^{(5)}_{A,\bar q}&=\frac{24}{5}\,K^A_{3\sigma},\\
\beta''^{(5)}_{A,q_\textrm{val}}&=\frac{-6}{5}\left(K^A_{\pi\pi}-2K^A_{33}-K^A_{\sigma\sigma}\right),&
\beta''^{(5)}_{A,q_\textrm{sea}}&=\frac{12}{5}\,K^A_{3\sigma},&
\beta''^{(5)}_{A,\bar q}&=\frac{-48}{5}\,K^A_{3\sigma}.
\end{align}
\end{subequations}
In the $7Q$ sector one has
\begin{subequations}
\begin{align}
\alpha''^{(7)}_{V,q_\textrm{val}}&=\frac{12}{5}\left(22K^V_{\pi\pi\pi\pi}+5K^V_{\pi\pi\pi\pi
2}+68K^V_{\sigma\sigma\pi\pi}+51K^V_{\sigma\sigma\sigma\sigma}\right),\\
\alpha''^{(7)}_{V,q_\textrm{sea}}&=\frac{12}{5}\left(17K^V_{\pi\pi\pi\pi}+2K^V_{\pi\pi\pi\pi
2}+42K^V_{\sigma\sigma\pi\pi}+27K^V_{\sigma\sigma\sigma\sigma}\right),\\
\alpha''^{(7)}_{V,\bar
q}&=\frac{-12}{5}\left(15K^V_{\pi\pi\pi\pi}-K^V_{\pi\pi\pi\pi
2}+46K^V_{\sigma\sigma\pi\pi}+30K^V_{\sigma\sigma\sigma\sigma}\right),\\
\beta''^{(7)}_{V,q_\textrm{val}}&=\frac{12}{5}\left(32K^V_{\pi\pi\pi\pi}+K^V_{\pi\pi\pi\pi
2}+88K^V_{\sigma\sigma\pi\pi}+57K^V_{\sigma\sigma\sigma\sigma}\right),\\
\beta''^{(7)}_{V,q_\textrm{sea}}&=\frac{12}{5}\left(19K^V_{\pi\pi\pi\pi}+2K^V_{\pi\pi\pi\pi
2}+62K^V_{\sigma\sigma\pi\pi}+45K^V_{\sigma\sigma\sigma\sigma}\right),\\
\beta''^{(7)}_{V,\bar
q}&=\frac{36}{5}\left(17K^V_{\pi\pi\pi\pi}+K^V_{\pi\pi\pi\pi
2}+50K^V_{\sigma\sigma\pi\pi}+34K^V_{\sigma\sigma\sigma\sigma}\right),
\end{align}
\end{subequations}
\begin{subequations}
\begin{align}
\alpha''^{(7)}_{A,q_\textrm{val}}&=\frac{12}{5}\left(3K^A_{\pi\pi\pi\pi
2}-2K^A_{\pi\pi 33}+10K^A_{\pi 3\pi
3}-10K^A_{\sigma\sigma\pi\pi}+34K^A_{\sigma\sigma
33}+19K^V_{\sigma\sigma\sigma\sigma}\right),\\
\alpha''^{(7)}_{A,q_\textrm{sea}}&=\frac{24}{5}\left(4K^A_{\pi\pi
3\sigma}+K^A_{\pi 3\pi\sigma}+13K^A_{\sigma\sigma
3\sigma}\right),\\
\alpha''^{(7)}_{A,\bar q}&=\frac{12}{5}\left(41K^A_{\pi\pi
3\sigma}-K^A_{\pi 3\pi\sigma}+80K^A_{\sigma\sigma
3\sigma}\right),\\
\beta''^{(7)}_{A,q_\textrm{val}}&=\frac{12}{5}\left(2K^A_{\pi\pi\pi\pi}-K^A_{\pi\pi\pi\pi
2}-18K^A_{\pi\pi 33}+26K^A_{\pi 3\pi
3}-22K^A_{\sigma\sigma\pi\pi}+50K^A_{\sigma\sigma
33}+17K^V_{\sigma\sigma\sigma\sigma}\right),\\
\beta''^{(7)}_{A,q_\textrm{sea}}&=\frac{24}{5}\left(10K^A_{\pi\pi
3\sigma}+K^A_{\pi 3\pi\sigma}+19K^A_{\sigma\sigma
3\sigma}\right),\\
\beta''^{(7)}_{A,\bar q}&=\frac{-36}{5}\left(23K^A_{\pi\pi
3\sigma}+K^A_{\pi 3\pi\sigma}+48K^A_{\sigma\sigma 3\sigma}\right).
\end{align}
\end{subequations}

One can easily check that the obvious sum rules for $\Theta^+$
\begin{subequations}
\begin{align}
\int\ud x\,[u(x)-\bar u(x)]&=2,\\
\int\ud x\,[d(x)-\bar d(x)]&=2,\\
\int\ud x\,[s(x)-\bar s(x)]&=-1
\end{align}
\end{subequations}
are satisfied separately in each sector. They are translated in our
parametrization as follows
\begin{subequations}
\begin{align}
\alpha''^{(i)}_{V,q_\textrm{val}}+\alpha''^{(i)}_{V,q_\textrm{sea}}-\alpha''^{(i)}_{V,\bar
q}&=\uN^{(i)}(B_{\bf\overline{10}}),\\
\beta''^{(i)}_{V,q_\textrm{val}}+\beta''^{(i)}_{V,q_\textrm{sea}}-\beta''^{(i)}_{V,\bar
q}&=0
\end{align}
\end{subequations}
for any $i=3Q,5Q,7Q,\cdots$.

A very interesting question about the pentaquark is its width. In
this model it is predicted to be very small (a few MeV) and can even
be $\lesssim 1$ MeV \cite{TimGhil}, quite unusual for baryons. In
the present approach this can be understood by the fact that since
there is no $3Q$ in the pentaquark and that in the Drell frame only
diagonal transitions in the Fock space occur, the decay is dominated
by the transition from the pentaquark $5Q$ sector to the proton $5Q$
sector, the latter being of course not so large. Since the
pentaquark production mechanism is not known, its width is estimated
by means of the the axial decay constant $\Theta^+\to K^+n$. If we
assume the approximate $SU(3)$ chiral symmetry one can obtain the
$\Theta\to KN$ pseudoscalar coupling from the generalized
Goldberger-Treiman relation
\begin{equation}
g_{\Theta KN}=\frac{g_A(\Theta\to KN)(M_\Theta+M_N)}{2F_K},
\end{equation}
where we use $M_\Theta=1530$ MeV, $M_N=940$ MeV and
$F_K=1.2F_\pi=112$ MeV. Once this transition pseudoscalar constant
is known, one can evaluate the $\Theta^+$ width from the general
expression for the $\frac{1}{2}^+$ hyperon decay \cite{Width}
\begin{equation}
\Gamma_\Theta=2\,\frac{g^2_{\Theta
KN}|\up|}{8\pi}\frac{(M_\Theta-M_N)^2-m_K^2}{M_\Theta^2},
\end{equation}
where
$|\up|=\sqrt{(M_\Theta^2-M_N^2-m_K^2)^2-4M_N^2m_K^2}/2M_\Theta=254$
MeV is the kaon momentum in the decay ($m_K=495$ MeV) and the factor
of 2 stands for the equal probability $K^+n$ and $K^0p$ decays.

Here are the combinations arising for this axial decay constant in
the $5Q$ and $7Q$ sectors
\begin{subequations}
\begin{align}
A^{(5)}(\Theta^+\to
K^+n)&=\frac{-6}{5}\sqrt{\frac{3}{5}}\left(7K^A_{\pi\pi}-8K^A_{33}+5K^A_{\sigma\sigma}-28K^A_{3\sigma}\right),\\
A^{(7)}(\Theta^+\to
K^+n)&=\frac{-48}{5}\sqrt{\frac{3}{5}}\left(7K^A_{\pi\pi\pi\pi}+7K^A_{\pi\pi\pi\pi
2}+6K^A_{\pi\pi 33}-14K^A_{\pi 3\pi
3}+40K^A_{\sigma\sigma\pi\pi}\right.\nonumber\\
&\quad\left.-38K^A_{\sigma\sigma
38}+22K^A_{\sigma\sigma\sigma\sigma}-71K^A_{\pi\pi 3\sigma}+K^A_{\pi
3\pi\sigma}-140K^A_{\sigma\sigma 3\sigma}\right).
\end{align}
\end{subequations}

\subsection{Numerical results}

In the evaluation of the scalar integrals we have used the
constituent quark mass $M=345$ MeV, the Pauli-Villars mass
$M_\textrm{PV}=556.8$ MeV for the regularization of \eqref{Direct}
and of \eqref{int} and the baryon mass $\uM=1207$ MeV as it follows
for the ``classical'' mass in the mean field approximation
\cite{Approximation}. The details of the computation are the same as
in \cite{Moi} where by choosing $\Phi^V(0,0)=1$ we had obtained in
the $3Q$ sector
\begin{equation}
\Phi^A(0,0)=0.8612
\end{equation}
and in the $5Q$ sector
\begin{subequations}
\begin{eqnarray}
&K^V_{\pi\pi}=0.03652,\qquad K^V_{33}=0.01975,\qquad
K^V_{\sigma\sigma}=0.01401,&
\end{eqnarray}
\begin{eqnarray}
&K^A_{\pi\pi}=0.03003,\qquad K^A_{33}=0.01628,\qquad
K^A_{\sigma\sigma}=0.01121,\qquad K^A_{3\sigma}=0.01626.&
\end{eqnarray}
\end{subequations}
Now come our results for the $7Q$ sector
\begin{subequations}
\begin{align}
K^V_{\pi\pi\pi\pi}&=0.00082, &K^V_{\pi\pi\pi\pi 2}&=0.00026,
&K^V_{\pi\pi 33}&=0.00039, &K^V_{3333}&=0.00019,\\
K^V_{\pi 3\pi 3}&=0.00017, &K^V_{\sigma\sigma\pi\pi}&=0.00027,
&K^V_{\sigma\sigma 33}&=0.00012,
&K^V_{\sigma\sigma\sigma\sigma}&=0.00009,
\end{align}
\begin{align}
K^A_{\pi\pi\pi\pi}&=0.00066, &K^A_{\pi\pi\pi\pi 2}&=0.00021,
&K^A_{\pi\pi 33}&=0.00031, &K^A_{3333}&=0.00015,\\
K^A_{\pi 3\pi 3}&=0.00013, &K^A_{\sigma\sigma\pi\pi}&=0.00021, &K^A_{\sigma\sigma 33}&=0.00010, &K^A_{\sigma\sigma\sigma\sigma}&=0.00007,\\
K^A_{\pi\pi 3\sigma}&=0.00031, &K^A_{333\sigma}&=0.00014, &K^A_{\pi
3\pi\sigma}&=0.00011, &K^A_{\sigma\sigma 3\sigma}&=0.00010.
\end{align}
\end{subequations}

The model has an intrinsic cutoff which is the instanton size $\sim
600$ MeV yielding the model scale $Q^2_0=0.36$ GeV$^2$.

\subsection{Discussion}

Let us start the discussion with our results for the normalizations.
They allow us to estimate which fraction of the proton is actually
made of $3Q$, $5Q$ and $7Q$. Since we did not compute the $7Q$
sector of decuplet baryons let us compare first the composition of
octet and decuplet baryons up to the $5Q$ sector.
\begin{table}[h!]\begin{center}\caption{\small{Comparison of octet and decuplet baryons fractions up to the $5Q$ sector.\newline}}
\begin{tabular}{c|cc} \hline\hline
&$3Q\equiv
\frac{\uN^{(3)}(B)}{\rule{0pt}{1.6ex}\uN^{(3)}(B)+\uN^{(5)}(B)}$&$5Q\equiv\frac{\uN^{(5)}(B)}{\rule{0pt}{1.6ex}\uN^{(3)}(B)+\uN^{(5)}(B)}$\rule[-2ex]{0pt}{5.5ex}\\\hline
\rule{0pt}{3ex} $B_{\bf 8}$&$77.5\%$&$22.5\%$\\\rule{0pt}{3ex}
$B_{\mathbf{10},3/2}$&$75\%$&$25\%$\\\rule{0pt}{3ex}
$B_{\mathbf{10},1/2}$&$72.5\%$&$27.5\%$\rule[-2ex]{0pt}{5ex}\\\hline\hline
\end{tabular}\label{pourcent1}\end{center}
\end{table}
\begin{table}[h!]\begin{center}
\caption{\small{Comparison of octet and antidecuplet baryons
fractions up to the $7Q$ sector.\newline}}
\begin{tabular}{c|ccc}
\hline\hline &$3Q\equiv
\frac{\uN^{(3)}(B)}{\rule{0pt}{1.6ex}\uN^{(3)}(B)+\uN^{(5)}(B)+\uN^{(7)}(B)}$&$5Q\equiv\frac{\uN^{(5)}(B)}{\rule{0pt}{1.6ex}\uN^{(3)}(B)+\uN^{(5)}(B)+\uN^{(7)}(B)}$&$7Q\equiv\frac{\uN^{(7)}(B)}{\rule{0pt}{1.6ex}\uN^{(3)}(B)+\uN^{(5)}(B)+\uN^{(7)}(B)}$\rule[-2ex]{0pt}{5.5ex}\\\hline
\rule{0pt}{3ex}
$B_{\mathbf{8}}$&$71.7\%$&$20.8\%$&$7.5\%$\\\rule{0pt}{3ex}
$B_{\mathbf{\overline{10}}}$&$0\%$&$60.6\%$&$39.4\%$\rule[-2ex]{0pt}{5ex}\\
\hline\hline
\end{tabular}\label{pourcent2}\end{center}
\end{table}

From Table \ref{pourcent1} one notices that the fractions are
similar for octet and decuplet baryons. The latter have a slightly
larger $5Q$ component, especially those with $J_z=1/2$. The fact
that decuplet baryons with $J_z=1/2$ and $J_z=3/2$ have different
composition is naturally related to a deviation of their shape from
sphericity, see previous discussion in subsection \ref{shape}.

From Table \ref{pourcent2} one observes that the dominant component
in pentaquarks is smaller ($\sim 60\%$) than the dominant one in
ordinary baryons ($\sim 75\%$). This would indicate that when
considering a pentaquark one should care more about higher Fock
contributions than in ordinary baryons. The additional
quark-antiquark pairs seem to be important to study exotic baryons.

Altogether, Tables \ref{pourcent1} and \ref{pourcent2} indicate that
roughly one fifth of the proton is actually made of $5Q$. This
result obtained without any fitting procedure is consistent with
estimations from other approaches, see \emph{e.g.}
\cite{fraction5q}.

We now proceed with our results for baryon vector and axial content.

\subsubsection{Octet content}

In Table \ref{Octetresult1} one can find the proton vector and axial
content.
\begin{table}[h!]\begin{center}\caption{\small{Our vector and axial content of the proton compared with NQM.\newline}}
\begin{tabular}{c|ccc|ccc|ccc}
\hline\hline
Vector&\multicolumn{3}{c|}{$u$}&\multicolumn{3}{c|}{$d$}&\multicolumn{3}{c}{$s$}\rule{0pt}{3ex}\\
&$\bar q$&$q_\textrm{sea}$&$q_\textrm{val}$&$\bar
q$&$q_\textrm{sea}$&$q_\textrm{val}$&$\bar
q$&$q_\textrm{sea}$&$q_\textrm{val}$\\\hline \rule{0pt}{3ex}
NQM&0&0&2&0&0&1&0&0&0\\\rule{0pt}{3ex}
$3Q$&0&0&2&0&0&1&0&0&0\\\rule{0pt}{3ex}
$3Q+5Q$&0.078&0.130&1.948&0.091&0.080&1.012&0.055&0.015&0.040\\\rule{0pt}{3ex}
$3Q+5Q+7Q$&0.125&0.202&1.924&0.145&0.128&1.017&0.088&0.028&0.060\rule[-2ex]{0pt}{5ex}\\
\hline
Axial&\multicolumn{3}{c|}{$\Delta u$}&\multicolumn{3}{c|}{$\Delta d$}&\multicolumn{3}{c}{$\Delta s$}\rule{0pt}{3ex}\\
&$\bar q$&$q_\textrm{sea}$&$q_\textrm{val}$&$\bar
q$&$q_\textrm{sea}$&$q_\textrm{val}$&$\bar
q$&$q_\textrm{sea}$&$q_\textrm{val}$\\\hline \rule{0pt}{3ex}
NQM&0&0&4/3&0&0&-1/3&0&0&0\\\rule{0pt}{3ex}
$3Q$&0&0&1.148&0&0&-0.287&0&0&0\\\rule{0pt}{3ex}
$3Q+5Q$&-0.032&-0.042&1.086&0.017&0.028&-0.275&0.005&0.005&-0.003\\\rule{0pt}{3ex}
$3Q+5Q+7Q$&-0.046&-0.060&1.056&0.026&0.040&-0.273&0.007&0.007&-0.006\rule[-2ex]{0pt}{5ex}\\
\hline\hline
\end{tabular}\label{Octetresult1}\end{center}
\end{table}
One can see that the sea is not $SU(3)$ symmetric ($\Delta\bar
u=\Delta\bar d=\Delta s=\Delta\bar s$) as naively often assumed. As
discussed earlier, this is due to the fact that we have a
non-perturbative sea of quark-antiquark pairs.
\newline

\textbf{Isospin asymmetry of the sea}
\newline

On the experimental side three collaborations SMC\cite{SMC}, HERMES
\cite{HERMES} and COMPASS \cite{COMPASS} have already measured
valence quark helicity distributions. In order to compare with our
results let us remind the relation between our ($\Delta q_{val}$)
and their definition of valence contribution ($\Delta q_v$)
\begin{equation}
\Delta q_v\equiv \Delta q_\textrm{val}+\Delta
q_\textrm{sea}-\Delta\bar q.
\end{equation}
Experiments favor an asymmetric light sea scenario $\Delta\bar
u=-\Delta\bar d$. Our results show indeed that $\Delta\bar u$ and
$\Delta\bar d$ have opposite signs but the contribution of
$\Delta\bar u$ is roughly twice the contribution of $\Delta\bar d$.
Concerning the sum $\Delta\bar u+\Delta\bar d$ it is about $2\%$
experimentally and is compatible with zero. The sum we have obtained
has the same order of magnitude but has the opposite sign. The DNS
parametrization gives $\Delta\bar u>0$ and $\Delta\bar d<0$ while
the statistical model \cite{Statistical} suggests the opposite signs
like our results. For the valence contribution, experiments suggest
$\Delta u_v+\Delta d_v\approx 0.40$ while we have obtained $\approx
0.76$.

Violation of Gottfried sum rule allows one to study also the vector
content of the sea. Experiments suggest that the $\bar d$ is
dominant over $\bar u$. This can physically be understood by
considering some simple Pauli-blocking effect. Since there are
already two valence $u$ quarks and only one valence $d$ quark in the
proton, the presence of $\bar d d$ pair will be favored compared to
$\bar u u$. The E866 collaboration \cite{Gottfried} gives $\bar
d-\bar u=0.118\pm 0.012$ while we have obtained $\bar d-\bar
u=0.019$. We indeed confirm an excess of $\bar d$ over $\bar u$ but
the magnitude is one order of magnitude too small.
\newline

\textbf{Strangeness contribution}
\newline

In Table \ref{Octetresult2} one can find the proton axial charges
and the flavor contributions to the proton spin compared with
experimental data.
\begin{table}[h!]\begin{center}\caption{\small{Our flavor contributions to the proton spin and axial charges compared with NQM and experimental data.\newline}}
\begin{tabular}{c|cccccc}
\hline\hline &$\Delta u$&$\Delta d$&$\Delta
s$&$g_A^{(3)}$&$g_A^{(8)}$&$g_A^{(0)}$\rule{0pt}{3ex}\\\hline
\rule{0pt}{3ex} NQM&4/3&-1/3&0&5/3&$1/\sqrt{3}$&1\\\rule{0pt}{3ex}
$3Q$&1.148&-0.287&0&1.435&0.497&0.861\\\rule{0pt}{3ex}
$3Q+5Q$&1.011&-0.230&0.006&1.241&0.444&0.787\\\rule{0pt}{3ex}
$3Q+5Q+7Q$&0.949&-0.207&0.009&1.156&0.419&0.751\\\rule{0pt}{3ex}
Exp. value \cite{PDG}&$0.83\pm 0.03$&$-0.43\pm 0.04$&$-0.10\pm 0.03$&$1.257\pm 0.003$&$0.34\pm 0.02$&$0.31\pm 0.07$\rule[-2ex]{0pt}{5ex}\\
\hline\hline
\end{tabular}\label{Octetresult2}\end{center}
\end{table}

Let us first concentrate on the strangeness contribution. We have
found a non-vanishing contribution $\Delta s$ which then naturally
breaks the Ellis-Jaffe sum rule. However compared to
phenomenological extractions \cite{Ds} it has the wrong sign and is
one order of magnitude too small. Even though there seems to be some
discrepancies among the extraction of $\Delta s$ by the different
methods, the strangeness contribution to proton spin is most likely
\emph{sizeable} and \emph{negative}. The approach we used is based
on flavor $SU(3)$ symmetry and we should, in fact, not expect to
obtain good quantitative results.

If we now have a look to the axial charges, even though the
individual flavor contribution are not satisfactory, we reproduce
fairly well $g_A^{(3)}$ without any fitting to the experimental
axial data. This is probably due to the fact that this isovector
axial charge is based on isospin $SU(2)$ symmetry and not on flavor
$SU(3)$. On the contrary, $g_A^{(8)}$ and $g_A^{(0)}$ extraction are
based on flavor $SU(3)$ symmetry. Even though we obtain that both
quark orbital angular momentum and quark-antiquark pairs reduce
their value compared with the NQM expectation, they are still far
too large, especially the isosinglet combination. Nevertheless, let
us remind that in the usual approach to $\chi$QSM, $g_A^{(0)}$ is
known to be sensitive to the strange quark mass $m_s$. It has been
shown that the latter reduces the fraction of spin carried by quarks
\cite{mscorr}. However, it has been recently argued that the
standard quantization scheme in chiral soliton models does not take
into account all necessary subleading contributions which are
essential when one is interested in strangeness issues \cite{Cohen}.

Note also that, as indicated by the $7Q$ component, one can
reasonably expect that adding further quark-antiquark pairs would
reduce further the axial charges but this reduction should be less
than 1\%.
\newline

\textbf{Axial decay constants}
\newline

In Table \ref{Octetresult3} one can find our results for octet axial
decay constants compared with the experimental knowledge. They are
in fair agreement. This is a nice result since, as we already
mentioned, it has been obtained without any fit to the corresponding
experimental data.
\begin{table}[h!]\begin{center}\caption{\small{Comparison of our octet axial decay constants with NQM predictions and experimental data.\newline}}
\begin{tabular}{c|cccc|c}
\hline\hline &NQM&$3Q$&$3Q+5Q$&$3Q+5Q+7Q$&Exp. value
\cite{PDG}\rule{0pt}{3ex}\\\hline \rule{0pt}{3ex}
$(g_A/g_V)_{n^0_\mathbf{ 8}\to
p^+_\mathbf{8}}$&5/3&1.435&1.241&1.156&$1.2695\pm
0.0029$\\\rule{0pt}{3ex} $(g_A/g_V)_{\Sigma^-_\mathbf{ 8}\to
\Sigma^0_\mathbf{8}}$&2/3&0.574&0.503&0.470&-\\\rule{0pt}{3ex}
$(g_A)_{\Sigma^-_\mathbf{ 8}\to
\Lambda^0_\mathbf{8}}$&$\sqrt{2/3}$&0.703&0.603&0.560&-\\\rule{0pt}{3ex}
$(g_A/g_V)_{\Sigma^0_\mathbf{ 8}\to
\Sigma^+_\mathbf{8}}$&2/3&0.574&0.503&0.470&-\\\rule{0pt}{3ex}
$(g_A)_{\Lambda^0_\mathbf{ 8}\to
\Sigma^+_\mathbf{8}}$&$\sqrt{2/3}$&0.703&0.603&0.560&-\\\rule{0pt}{3ex}
$(g_A/g_V)_{\Xi^-_\mathbf{ 8}\to
\Xi^0_\mathbf{8}}$&-1/3&-0.287&-0.236&-0.215&-\\\rule{0pt}{3ex}
$(g_A/g_V)_{\Sigma^-_\mathbf{ 8}\to
n^0_\mathbf{8}}$&-1/3&-0.287&-0.236&-0.215&$-0.340\pm
0.017$\\\rule{0pt}{3ex} $(g_A/g_V)_{\Xi^-_\mathbf{ 8}\to
\Sigma^0_\mathbf{8}}$&5/3&1.435&1.241&1.156&-\\\rule{0pt}{3ex}
$(g_A/g_V)_{\Xi^-_\mathbf{ 8}\to
\Lambda^0_\mathbf{8}}$&1/3&0.287&0.256&0.242&$0.25\pm
0.05$\\\rule{0pt}{3ex} $(g_A/g_V)_{\Sigma^0_\mathbf{ 8}\to
p^+_\mathbf{8}}$&-1/3&-0.287&-0.236&-0.215&-\\\rule{0pt}{3ex}
$(g_A/g_V)_{\Lambda^0_\mathbf{ 8}\to
p^+_\mathbf{8}}$&1&0.861&0.749&0.699&$0.718\pm
0.015$\\\rule{0pt}{3ex} $(g_A/g_V)_{\Xi^0_\mathbf{ 8}\to
\Sigma^+_\mathbf{8}}$&5/3&1.435&1.241&1.156&$1.21\pm 0.05$\rule[-2ex]{0pt}{5ex}\\
\hline\hline
\end{tabular}\label{Octetresult3}\end{center}
\end{table}

In the literature, these octet axial transitions are often described
in terms of the $F\&D$ parametrization. Under the assumption of
flavor $SU(3)$ symmetry, the $F\&D$ parameters can be extracted from
the experimentally known axial constants. These parameters are
compared with our values for $F\&D$ in Table \ref{Octetresult4}. One
can see that our values are closer to experimental values than the
expectation of NQM. In particular $F$ is well reproduced but $D$ is
too small.
\begin{table}[h!]\begin{center}\caption{\small{Comparison of our $F\& D$ parameters with NQM predictions and $SU(3)$ fits to experimental data.\newline}}
\begin{tabular}{c|cccc|c}
\hline\hline &NQM&$3Q$&$3Q+5Q$&$3Q+5Q+7Q$&$SU(3)$ fit \cite{FitSU3}
\rule{0pt}{3ex}\\\hline \rule{0pt}{3ex}
$F$&2/3&0.574&0.503&0.470&$0.475\pm 0.004$\\\rule{0pt}{3ex}
$D$&1&0.861&0.739&0.686&$0.793\pm 0.005$\\\rule{0pt}{3ex}
$F/D$&2/3&2/3&0.680&0.686&$0.599\pm 0.006$\\\rule{0pt}{3ex}
$3F-D$&1&0.861&0.769&0.725&$0.632\pm 0.017$\rule[-2ex]{0pt}{5ex}\\
\hline\hline
\end{tabular}\label{Octetresult4}\end{center}
\end{table}
\newpage

\subsubsection{Decuplet content}

In Tables \ref{Decupletresult3/2} and \ref{Decupletresult1/2} one
can find the $\Delta^{++}$ vector and axial content with
respectively $J_z=3/2,1/2$.
\begin{table}[h!]\begin{center}\caption{\small{Our vector and axial content of the $\Delta^{++}$ with spin projection $J_z=3/2$ compared with NQM.\newline}}
\begin{tabular}{c|ccc|ccc|ccc}
\hline\hline
Vector&\multicolumn{3}{c|}{$u$}&\multicolumn{3}{c|}{$d$}&\multicolumn{3}{c}{$s$}\rule{0pt}{3ex}\\
$J_z=3/2$&$\bar q$&$q_\textrm{sea}$&$q_\textrm{val}$&$\bar
q$&$q_\textrm{sea}$&$q_\textrm{val}$&$\bar
q$&$q_\textrm{sea}$&$q_\textrm{val}$\\\hline \rule{0pt}{3ex}
NQM&0&0&3&0&0&0&0&0&0\\\rule{0pt}{3ex}
$3Q$&0&0&3&0&0&0&0&0&0\\\rule{0pt}{3ex}
$3Q+5Q$&0.072&0.193&2.879&0.089&0.029&0.060&0.089&0.029&0.060\rule[-2ex]{0pt}{5ex}\\
\hline
Axial&\multicolumn{3}{c|}{$\Delta u$}&\multicolumn{3}{c|}{$\Delta d$}&\multicolumn{3}{c}{$\Delta s$}\rule{0pt}{3ex}\\
$J_z=3/2$&$\bar q$&$q_\textrm{sea}$&$q_\textrm{val}$&$\bar
q$&$q_\textrm{sea}$&$q_\textrm{val}$&$\bar
q$&$q_\textrm{sea}$&$q_\textrm{val}$\\\hline \rule{0pt}{3ex}
NQM&0&0&3&0&0&0&0&0&0\\\rule{0pt}{3ex}
$3Q$&0&0&2.538&0&0&0&0&0&0\\\rule{0pt}{3ex}
$3Q+5Q$&-0.061&-0.079&2.423&0.012&0.021&-0.015&0.012&0.021&-0.015\rule[-2ex]{0pt}{5ex}\\
\hline\hline
\end{tabular}\label{Decupletresult3/2}\end{center}
\end{table}
In Tables \ref{Result3/2} and \ref{Result1/2} one can find the
$\Delta^{++}$ axial charges with respectively $J_z=3/2,1/2$.
\begin{table}[h!]\begin{center}\caption{\small{Our vector and axial content of the $\Delta^{++}$ with spin projection $J_z=1/2$ compared with NQM.\newline}}
\begin{tabular}{c|ccc|ccc|ccc}
\hline\hline
Vector&\multicolumn{3}{c|}{$u$}&\multicolumn{3}{c|}{$d$}&\multicolumn{3}{c}{$s$}\rule{0pt}{3ex}\\
$J_z=1/2$&$\bar q$&$q_\textrm{sea}$&$q_\textrm{val}$&$\bar
q$&$q_\textrm{sea}$&$q_\textrm{val}$&$\bar
q$&$q_\textrm{sea}$&$q_\textrm{val}$\\\hline \rule{0pt}{3ex}
NQM&0&0&3&0&0&0&0&0&0\\\rule{0pt}{3ex}
$3Q$&0&0&3&0&0&0&0&0&0\\\rule{0pt}{3ex}
$3Q+5Q$&0.059&0.225&2.834&0.108&0.025&0.083&0.108&0.025&0.083\rule[-2ex]{0pt}{5ex}\\
\hline
Axial&\multicolumn{3}{c|}{$\Delta u$}&\multicolumn{3}{c|}{$\Delta d$}&\multicolumn{3}{c}{$\Delta s$}\rule{0pt}{3ex}\\
$J_z=1/2$&$\bar q$&$q_\textrm{sea}$&$q_\textrm{val}$&$\bar
q$&$q_\textrm{sea}$&$q_\textrm{val}$&$\bar
q$&$q_\textrm{sea}$&$q_\textrm{val}$\\\hline \rule{0pt}{3ex}
NQM&0&0&1&0&0&0&0&0&0\\\rule{0pt}{3ex}
$3Q$&0&0&0.861&0&0&0&0&0&0\\\rule{0pt}{3ex}
$3Q+5Q$&-0.020&-0.026&0.813&0.004&0.007&-0.007&0.004&0.007&-0.007\rule[-2ex]{0pt}{5ex}\\
\hline\hline
\end{tabular}\label{Decupletresult1/2}\end{center}
\end{table}

To the best of our knowledge there is no experimental results
concerning the vector and axial properties of decuplet baryons. Our
results can then be considered just as theoretical predictions, at
least qualitatively. We would like to emphasize a few observations:
\begin{itemize}
\item Like in the proton, quark spins alone do not add
up to the total decuplet baryon spin. The missing spin has to be
attributed to orbital angular momentum of quarks and additional
quark-antiquark pairs.
\item We have obtained that the polarization of the ``hidden'' flavors
$\Delta d$ and $\Delta s$ in $\Delta^{++}$ has the same sign as the
``hidden'' flavor $\Delta s$ in the proton.
\item In general, for the ``hidden'' flavor contributions, we have $\Delta
q_\textrm{sea}\neq\Delta\bar q$. The only exception is the octet
where $\Delta q_\textrm{sea}=\Delta\bar q$ for ``hidden'' flavor is
satisfied at the $5Q$ level, see eq. \eqref{octet5Q}. The $7Q$
contribution however satisfies $\Delta q_\textrm{sea}\neq\Delta\bar
q$, see eqs. \eqref{Octet7Q} and \eqref{Octet7Q2}, and so the
exception appears just as a mere coincidence. The numerical values
appearing in Table \ref{Octetresult1} at the $7Q$ level for $\Delta
s_\textrm{sea}$ and $\Delta\bar s$ are quite close and the
differences appear only in the fourth decimal $\Delta\bar s=0.0073$,
$\Delta s_\textrm{sea}=0.0075$.
\end{itemize}
\begin{table}[h!]\begin{center}\caption{\small{Flavor contributions to the $\Delta^{++}_{J_z=3/2}$ spin and axial charges compared with NQM.\newline}}
\begin{tabular}{c|cccccc}
\hline\hline &$\Delta u$&$\Delta d$&$\Delta
s$&$g_A^{(3)}$&$g_A^{(8)}$&$g_A^{(0)}$\rule{0pt}{3ex}\\\hline
\rule{0pt}{3ex} NQM&3&0&0&3&$\sqrt{3}$&3\\\rule{0pt}{3ex}
$3Q$&2.583&0&0&2.583&1.492&2.583\\\rule{0pt}{3ex}
$3Q+5Q$&2.283&0.018&0.018&2.265&1.307&2.319\rule[-2ex]{0pt}{5ex}\\
\hline\hline
\end{tabular}\label{Result3/2}\end{center}
\end{table}
\begin{table}[h!]\begin{center}\caption{\small{Flavor contributions to the $\Delta^{++}_{J_z=1/2}$ spin and axial charges compared with NQM.\newline}}
\begin{tabular}{c|cccccc}
\hline\hline &$\Delta u$&$\Delta d$&$\Delta
s$&$g_A^{(3)}$&$g_A^{(8)}$&$g_A^{(0)}$\rule{0pt}{3ex}\\\hline
\rule{0pt}{3ex} NQM&1&0&0&1&$1/\sqrt{3}$&1\\\rule{0pt}{3ex}
$3Q$&0.861&0&0&0.861&0.497&0.861\\\rule{0pt}{3ex}
$3Q+5Q$&0.767&0.004&0.004&0.763&0.441&0.775\rule[-2ex]{0pt}{5ex}\\
\hline\hline
\end{tabular}\label{Result1/2}\end{center}
\end{table}

\subsubsection{Antidecuplet content}

The study of the $7Q$ sector has mainly been motivated by the
pentaquark. In previous works \cite{Moi} we have shown that the $5Q$
component of usual baryons has non-negligible and interesting
effects on the vector and axial quantities. In the same spirit,
since there is no $3Q$ component in pentaquarks, it would be
interesting to see what happens when one considers the $7Q$
component. In Table \ref{Antidecupletresult} one can find the
$\Theta^+$ vector and axial content and in Table
\ref{Antidecupletresult2} the antidecuplet axial charges.
\begin{table}[h!]\begin{center}\caption{\small{Our vector and axial content of the $\Theta^+$.\newline}}
\begin{tabular}{c|ccc|ccc|ccc}
\hline\hline
Vector&\multicolumn{3}{c|}{$u$}&\multicolumn{3}{c|}{$d$}&\multicolumn{3}{c}{$s$}\rule{0pt}{3ex}\\
&$\bar q$&$q_\textrm{sea}$&$q_\textrm{val}$&$\bar
q$&$q_\textrm{sea}$&$q_\textrm{val}$&$\bar
q$&$q_\textrm{sea}$&$q_\textrm{val}$\\\hline \rule{0pt}{3ex}
$5Q$&0&1/2&3/2&0&1/2&3/2&1&0&0\\\rule{0pt}{3ex}
$5Q+7Q$&0.153&0.680&1.474&0.153&0.680&1.474&1.088&0.035&0.053\rule[-2ex]{0pt}{5ex}\\
\hline
Axial&\multicolumn{3}{c|}{$\Delta u$}&\multicolumn{3}{c|}{$\Delta d$}&\multicolumn{3}{c}{$\Delta s$}\rule{0pt}{3ex}\\
&$\bar q$&$q_\textrm{sea}$&$q_\textrm{val}$&$\bar
q$&$q_\textrm{sea}$&$q_\textrm{val}$&$\bar
q$&$q_\textrm{sea}$&$q_\textrm{val}$\\\hline \rule{0pt}{3ex}
$5Q$&0&0.322&0.136&0&0.322&0.136&0.644&0&0\\\rule{0pt}{3ex}
$5Q+7Q$&-0.020&0.276&0.113&-0.020&0.276&0.113&0.610&0.019&-0.014\rule[-2ex]{0pt}{5ex}\\
\hline\hline
\end{tabular}\label{Antidecupletresult}\end{center}
\end{table}
\begin{table}[h!]\begin{center}\caption{\small{Flavor contributions to the $\Theta^{+}$ spin and axial charges.\newline}}
\begin{tabular}{c|cccccc}
\hline\hline &$\Delta u$&$\Delta d$&$\Delta
s$&$g_A^{(3)}$&$g_A^{(8)}$&$g_A^{(0)}$\rule{0pt}{3ex}\\\hline
\rule{0pt}{3ex}
$5Q$&0.458&0.458&0.644&0&-0.215&1.560\\\rule{0pt}{3ex}
$5Q+7Q$&0.369&0.369&0.615&0&-0.284&1.353\rule[-2ex]{0pt}{5ex}\\
\hline\hline
\end{tabular}\label{Antidecupletresult2}\end{center}
\end{table}

The first interesting thing here is that contrarily to usual baryons
the sum of all quark spins is larger than the total baryon spin.
This means that quark spins are mainly parallel to the baryon spin
and that their orbital angular momentum is opposite in order to
compensate and form at the end a baryon with spin $1/2$.

The second interesting thing is that this $7Q$ component does not
change qualitatively the results given by the $5Q$ sector alone.
This means that a rather good estimation of pentaquark properties
can be obtained by means of the dominant sector only.

\begin{table}[h!]\begin{center}\caption{\small{$\Theta^{+}$
width estimation.\newline}}
\begin{tabular}{c|ccc}
\hline\hline
&$g_A(\Theta\to KN)$&$g_{\Theta KN}$&$\Gamma_\Theta$ (MeV)\rule{0pt}{3ex}\\
\hline\rule{0pt}{3ex} $5Q$&0.144&1.592&2.256\\\rule{0pt}{3ex}
$5Q+7Q$&0.169&1.864&3.091\rule[-2ex]{0pt}{5ex}\\
\hline\hline
\end{tabular}\label{Thetawidth}\end{center}
\end{table}

We close this discussion by looking at the width of $\Theta^+$
pentaquark, see Table \ref{Thetawidth}. The $7Q$ component does not
change much our previous estimation. Note however, as one could have
expected, that the width is slightly increased. Indeed it has been
explained in previous papers \cite{DiaPet,Moi} that the unusually
small width of pentaquarks can be understood in the present approach
by the fact that the pentaquark cannot decay into the $3Q$ sector of
the nucleon. Since the addition of a $7Q$ component reduces the
overall weight of the $3Q$ component in the nucleon (see Tables
\ref{pourcent1} and \ref{pourcent2}) the width should increase. The
existence of a narrow pentaquark resonance within $\chi$QSM is safe
and appears naturally without any parameter fixing.

\section{Conclusion}

The question of the nucleon structure is one of the most intriguing
in the field of strong interactions. Experiments indicate many
non-trivial effects that are related to the non-perturbative regime
of QCD. The question of identifying the relevant degrees of freedom
is still open and many models have been studied to understand and
accommodate experimental results. The Chiral-Quark Soliton Model
($\chi$QSM) is one of them and has already given lots of successful
results. Recently this model has been formulated in the Infinite
Momentum Frame (IMF) where it was possible to write down a general
expression for the wave function of any light baryon. Using this
unique tool, one could in principle access to a large amount of
information concerning the structure and the properties of the
nucleon at low energies.

In this paper we have presented our results concerning the octet,
decuplet and antidecuplet spin and flavor structure up to the $7Q$
Fock sector. The model being based on the collective quantization of
the solitonic pion field, an expression for the spin-flavor wave
function can be and has been obtained for the $3Q$, $5Q$ and $7Q$
sectors. In previous works it has been shown that the technique
reproduces the $SU(6)$ wave functions in the $3Q$ sector. Remarkably
the soliton \emph{Ansatz} allows one to extract an exact form for
higher Fock components without free coefficients and can serve as a
basis for other quark models which aim to include higher Fock
components.

Although our approach is restricted to flavor $SU(3)$ symmetry we
have obtained a fairly good description, especially concerning the
octet axial decay constants. Among the discrepancies let us note a
too large value for the quark spin contribution to the baryon spin
which could in principle be solved by considering the breakdown of
flavor $SU(3)$ symmetry. Compared to what is suggested by
experiments it seems that the contribution of our sea to the axial
charges is by an order of magnitude too small and has the opposite
sign. However the adjunction of additional quark-antiquark pairs and
quark orbital angular momentum brings the quantities closer to the
experimental values compared to the Naive Quark Model predictions.

Within this approach we have also shown explicitly that the decuplet
baryons are not spherical, due to the pion field. Indeed a
quadrupole structure naturally appeared already in the $5Q$ sector.
We have also obtained an interesting result concerning the
pentaquark spin. Contrarily to ordinary baryons, the sum of quark
spins is larger than the total pentaquark spin and thus that orbital
angular momentum is antiparallel to the total spin. This approach is
particularly interesting since one can easily distinguish between
valence quark, sea quark and antiquark contributions and therefore
allows one to study explicitly the sea.

In a previous work, we have shown that the $5Q$ component in
ordinary baryons is important especially to explain the spin
distributions. By analogy it was then interesting to study the
influence of the $7Q$ component in pentaquarks. Qualitatively the
results are not changed and thus justify \emph{a posteriori} the
validity of the expansion in the number of quark-antiquark pairs.
The particular feature of pentaquarks is their unusual small width.
In the present approach this smallness is explained by the fact that
the pentaquark cannot decay into the $3Q$ component of the nucleon.
Consequently one can expect that adding higher Fock states would
decrease $3Q$ component of the nucleon and thus increase the
pentaquark width. This pattern has indeed been obtained.

The results are given without any estimate of theoretical errors
since the latter are rather difficult to evaluate at the present
stage of the study. We emphasize also that in this work we did not
fit any parameter. The sole parameters of the model were fitted in
the meson sector.

\subsection*{Acknowledgements}

The author is grateful to RUB TP2 for its kind hospitality, to D.
Diakonov for suggesting this work and to M. Polyakov for his careful
reading and comments. The author is also indebted to J. Cugnon for
his help and advices. This work has been supported by the National
Funds of Scientific Research, Belgium.

\newpage
\renewcommand{\theequation}{A\arabic{equation}}
\setcounter{equation}{0}

\section*{Appendix A: Group integrals}

We give in this Appendix the complete list of octet, decuplet and
antidecuplet spin-flavor wave functions up to the $7Q$ sector. They
are group integrals over the Haar measure of the $SU(N)$ group which
is normalized to unity $\int\ud R=1$. Part of them are copied from
the Appendix B of \cite{DiaPet}.

\subsection*{A.1 Method}

Here is the general method to compute integrals of several matrices
$R$, $R^\dag$. The result of an integration over the invariant
measure can only be invariant tensors which, for the $SU(N)$ group,
are built solely from the Kronecker $\delta$ and Levi-Civita
$\epsilon$ tensors. One then constructs the supposed tensor of a
given rank as the most general combination of $\delta$'s and
$\epsilon$'s satisfying the symmetry relations following from the
integral in question:
\begin{itemize}\item Since
$R^f_j$ and $R^{\dag i}_h$ are just numbers one can commute them.
Therefore the same permutation among $f$'s and $j$'s (or $h$'s and
$i$'s) does not change the value of the integral, \emph{i.e.} the
structure of the tensor.\item In the special case where there are as
many $R$ as $R^\dag$, one can exchange them which amounts to
exchange $f$ and $j$ indices with respectively $i$ and $h$.
\end{itemize}
One has however to be careful to use the same ``type'' of indices in
$\delta$'s and $\epsilon$'s, \emph{i.e.} the upper (resp. lower)
indices of $R$ with the lower (resp. upper) ones of $R^\dag$. The
indefinite coefficients in the combination are found by contracting
both sides with various $\delta$'s and $\epsilon$'s and thus by
reducing the integral to a previously derived one. We will give
below explicit examples.

\subsection*{A.2 Basic integrals and explicit examples}

Since the method is recursive, let us start with the simplest group
integrals. For any $SU(N)$ group one has
\begin{equation}\label{Basic}
\int\ud R\,R^f_j=0,\qquad \int\ud R\,R^{\dag i}_h=0,\qquad \int\ud
R\, R^f_jR^{\dag i}_h=\frac{1}{N}\,\delta^f_h\delta^i_j.
\end{equation}
The last integral is a well known result but can be derived by means
of the method explained earlier. There are two upper ($f,i$) and two
lower ($j,h$) indices. In $SU(N)$ the solution of the integral can
only be constructed from the $\delta$ and the $\epsilon$ tensor with
$N$ (upper or lower) indices. There is only one possible
structure\footnote{The $\epsilon$ tensor needs $N$ indices of the
same ``type'' and position. The only possibility left is to
introduce new indices that are summed, \emph{e.g.}
$\epsilon^{fg}\epsilon_{hg}\epsilon^{ik}\epsilon_{jk}$. This is
however not a new structure since the summation over the new indices
can be performed leading to the ``old'' structure
$\epsilon^{fg}\epsilon_{hg}\epsilon^{ik}\epsilon_{jk}=\delta^f_h\delta^i_j$.}
$\delta^f_h\delta^i_j$ leaving thus only one undetermined
coefficient $A$. This coefficient can be determined by contracting
both sides with, say, $\delta^j_i$. Since $R^f_jR^{\dag
j}_h=\delta^f_h$ ($R$ matrices belong to $SU(N)$ and are thus
unitary) one has for the lhs
\begin{equation}
\delta^j_i\times\int\ud R\,R^f_jR^{\dag i}_h=\delta^f_h
\end{equation}
and for the rhs
\begin{equation}
\delta^j_i\times A\,\delta^f_h\delta^i_j=A\,N\,\delta^f_h
\end{equation}
and one concludes that $A=1/N$.

Let us proceed with the integral of two $R$'s. Here all the upper
(lower) indices have the same ``type'' and must appear in the same
symbol. Only $\epsilon$ has many indices in the same position. In
the case $N>2$ one needs more available indices. This means that for
$SU(N)$ with $N>2$ one has
\begin{equation}
\int\ud R\,R^{f_1}_{j_1}R^{f_2}_{j_2}=0.
\end{equation}
For $N=2$, the group integral is non-vanishing since the structure
$\epsilon^{f_1f_2}\epsilon_{j_1j_2}$ is allowed. The undetermined
coefficient $A$ is obtained by contracting both sides with, say,
$\epsilon^{j_1j_2}$. Since
$\epsilon^{j_1j_2}R^{f_1}_{j_1}R^{f_2}_{j_2}=\epsilon^{f_1f_2}$ ($R$
matrices belong to $SU(2)$ and have thus $\det(R)=1$) one has for
the lhs
\begin{equation}
\epsilon^{j_1j_2}\times\int\ud
R\,R^{f_1}_{j_1}R^{f_2}_{j_2}=\epsilon^{f_1f_2}
\end{equation}
and for the rhs
\begin{equation}
\epsilon^{j_1j_2}\times
A\,\epsilon^{f_1f_2}\epsilon_{j_1j_2}=2A\,\epsilon^{f_1f_2}
\end{equation}
and thus one concludes that $A=1/2$. For $SU(2)$ one then has
\begin{equation}
\int\ud
R\,R^{f_1}_{j_1}R^{f_2}_{j_2}=\frac{1}{2}\,\epsilon^{f_1f_2}\epsilon_{j_1j_2}.
\end{equation}

The $SU(3)$ analog involves the products of three $R$'s
\begin{equation}
\int\ud
R\,R^{f_1}_{j_1}R^{f_2}_{j_2}R^{f_3}_{j_3}=\frac{1}{6}\,\epsilon^{f_1f_2f_3}\epsilon_{j_1j_2j_3}
\end{equation}
which is vanishing for $N>3$ and also for $N=2$ since all the three
upper (and lower) indices cannot be used in $\epsilon$'s. This can
be easily generalized to $SU(N)$ with the product of $N$ matrices
$R$
\begin{equation}
\int\ud R\,R^{f_1}_{j_1}R^{f_2}_{j_2}\dots
R^{f_N}_{j_N}=\frac{1}{N!}\,\epsilon^{f_1f_2\dots
f_N}\epsilon_{j_1j_2\dots j_N}.
\end{equation}
This integral is vanishing for all $SU(N')$ groups with $N'$ that is
not a divisor of $N$.

Let us now consider the product of four $R$'s in $SU(2)$. Since 2 is
a divisor of 4 the integral is non-vanishing. The general tensor
structure is a linear combination of
$\epsilon^{f_af_b}\epsilon^{f_cf_d}\epsilon_{j_wj_x}\epsilon_{j_yj_z}$
with $a,b,c,d$ and $w,x,y,z$ some permutation of 1,2,3,4. There are
\emph{a priori} 9 undetermined coefficients. The symmetries of the
integral reduce this number to 2. Thanks to the $SU(2)$ identity
\begin{equation}\label{Simplification}
\epsilon_{j_1j_2}\epsilon_{j_3j_4}+\epsilon_{j_1j_3}\epsilon_{j_4j_2}+\epsilon_{j_1j_4}\epsilon_{j_2j_3}=0
\end{equation}
only one undetermined coefficient is left which is obtained by
contracting both sides with, say, $\epsilon^{j_1j_2}$. The result is
thus for $SU(2)$
\begin{equation}
\int\ud
R\,R^{f_1}_{j_1}R^{f_2}_{j_2}R^{f_3}_{j_3}R^{f_4}_{j_4}=\frac{1}{6}\,\left(\epsilon^{f_1f_2}\epsilon^{f_3f_4}\epsilon_{j_1j_2}\epsilon_{j_3j_4}+\epsilon^{f_1f_3}\epsilon^{f_2f_4}\epsilon_{j_1j_3}\epsilon_{j_2j_4}+\epsilon^{f_1f_4}\epsilon^{f_2f_3}\epsilon_{j_1j_4}\epsilon_{j_2j_3}\right).
\end{equation}

The identity (\ref{Simplification}) is in fact a particular case of
a general $SU(N)$ identity. It is based on the fact that for $SU(N)$
one has $\epsilon_{j_1j_2\dots j_{N+1}}=0$ and thus
\begin{equation}\label{Formule}
\epsilon_{j_1j_2\dots j_N}X_{j_{N+1}}\pm\epsilon_{j_2j_3\dots
j_{N+1}}X_{j_1}+\epsilon_{j_3j_4\dots
j_1}X_{j_2}\pm\ldots\pm\epsilon_{j_{N+1}j_1\dots j_{N-1}}X_{j_N}=0
\end{equation}
where the $+$ (resp. $-$) sign is for $N$ even (resp. odd) and $X_j$
any tensor with at least index $j$. This identity is easy to check.
Since we work in $SU(N)$, among the $N+1$ indices at least two are
equal, say $j_k$ and $j_l$. The only surviving terms are then
$-X_{j_k}+X_{j_l}$ which give zero since $j_k=j_l$. It is very
useful and greatly simplifies the search of the general tensor
structure. Since the number of indices of both ``types'' is
identical, the structure in terms of $\delta$'s and $\epsilon$'s is
also the same. The indices on $\epsilon$ can be placed in a
symmetric (\emph{e.g.}
$\epsilon^{f_1f_2}\epsilon^{f_3f_4}\epsilon_{j_1j_2}\epsilon_{j_3j_4}$)
and an asymmetric manner (\emph{e.g.}
$\epsilon^{f_1f_2}\epsilon^{f_3f_4}\epsilon_{j_1j_4}\epsilon_{j_2j_3}$).
By repeated applications of (\ref{Formule}) the asymmetric part of
the tensor can be transformed into the symmetric part reducing thus
the number of undetermined coefficients by a factor 2. In the search
of the general tensor structure one can then just consider symmetric
$\epsilon$ terms only.

We give another useful identity. In $SU(2)$ one has
$\epsilon^{f_1f_2f_3}\epsilon_{h_1h_2h_3}=0$. Using the notation
$(abc)\equiv\delta^{f_1}_{h_a}\delta^{f_2}_{h_b}\delta^{f_3}_{h_c}$
this amounts to
\begin{equation}\label{Formule2}
(123)-(132)+(231)-(213)+(312)-(321)=0.
\end{equation}
This identity is easily generalized to any $SU(N)$ group where it is
based on $\epsilon^{f_1f_2\dots f_{N+1}}\epsilon_{h_1h_2\dots
h_{N+1}}=0$.

We close this section by mentioning another group integral which is
useful to obtain further ones. For any $SU(N)$ group one has
\begin{equation}
\int\ud R\,R^{f_1}_{j_1}R^{f_2}_{j_2}R^{\dag i_1}_{h_1}R^{\dag
i_2}_{h_2}=\frac{1}{N^2-1}\left[\delta^{f_1}_{h_1}\delta^{f_2}_{h_2}\left(\delta^{i_1}_{j_1}\delta^{i_2}_{j_2}-\frac{1}{N}\,\delta^{i_2}_{j_1}\delta^{i_1}_{j_2}\right)
+\delta^{f_1}_{h_2}\delta^{f_2}_{h_1}\left(\delta^{i_2}_{j_1}\delta^{i_1}_{j_2}-\frac{1}{N}\,\delta^{i_1}_{j_1}\delta^{i_2}_{j_2}\right)\right].
\end{equation}
One can easily check that by contracting it with, say,
$\delta^{h_1}_{f_1}$ it reduces to (\ref{Basic}).

\subsection*{A.3 Notations}

In order to simplify the formulae we introduce a few notations
\begin{equation}
\left[abc\right]\equiv(123)(abc)+(231)(bca)+(312)(cab)+(213)(bac)+(132)(acb)+(321)(cba),
\end{equation}
\begin{equation}
\begin{split}
\left[abcd\right]\equiv&(1234)(abcd)+(2341)(bcda)+(3412)(cdab)+(4123)(dabc)+(2134)(bacd)+(1342)(acdb)\\
&+(3421)(cdba)+(4213)(dbac)+(3214)(cbad)+(2143)(badc)+(1432)(adcb)+(4321)(dcba)\\
&+(4231)(dbca)+(2314)(bcad)+(3142)(cadb)+(1423)(adbc)+(1324)(acbd)+(3241)(cbda)\\
&+(2413)(bdac)+(4132)(dacb)+(1243)(abdc)+(2431)(bdca)+(4312)(dcab)+(3124)(cabd),
\end{split}
\end{equation}
\begin{equation}
\begin{split}
\left[abcde\right]\equiv&(12345)(abcde)+(23451)(bcdea)+(34512)(cdeab)+(45123)(deabc)+(51234)(eabcd)\\
&+(21345)(bacde)+(13452)(acdeb)+(34521)(cdeba)+(45213)(debac)+(52134)(ebacd)\\
&+(32145)(cbade)+(21453)(badec)+(14532)(adecb)+(45321)(decba)+(53214)(ecbad)\\
&+(42315)(dbcae)+(23154)(bcaed)+(31542)(caedb)+(15423)(aedbc)+(54231)(edbca)\\
&+(52341)(ebcda)+(23415)(bcdae)+(34152)(cdaeb)+(41523)(daebc)+(15234)(aebcd)\\
&+(13245)(acbde)+(32451)(cbdea)+(24513)(bdeac)+(45132)(deacb)+(51324)(eacbd)\\
&+(14325)(adcbe)+(43251)(dcbea)+(32514)(cbead)+(25143)(beadc)+(51432)(eadcb)\\
&+(15342)(aecdb)+(53421)(ecdba)+(34215)(cdbae)+(42153)(dbaec)+(21534)(baecd)\\
&+(12435)(abdce)+(24351)(bdcea)+(43512)(dceab)+(35124)(ceabd)+(51243)(eabdc)\\
&+(12543)(abedc)+(25431)(bedca)+(54312)(edcab)+(43125)(dcabe)+(31254)(cabed)\\
&+(12354)(abced)+(23541)(bcdea)+(35412)(cedab)+(54123)(edabc)+(41235)(dabce)\\
&+(54321)(edcba)+(43215)(dcbae)+(32154)(cbaed)+(21543)(baedc)+(15432)(aedcb)\\
&+(12453)(abdec)+(24531)(bdeca)+(45312)(decab)+(53124)(ecabd)+(31245)(cabde)\\
&+(12534)(abecd)+(25341)(becda)+(53412)(ecdab)+(34125)(cdabe)+(41253)(dabec)\\
&+(23514)(bcead)+(35142)(ceadb)+(51423)(eadbc)+(14235)(adbce)+(42351)(dbcea)\\
&+(23145)(bcade)+(31452)(cadeb)+(14523)(adebc)+(45231)(debca)+(52314)(ebcad)\\
&+(34251)(cdbea)+(42513)(dbeac)+(25134)(beacd)+(51342)(eacdb)+(13425)(acdbe)\\
&+(21435)(badce)+(14352)(adceb)+(43521)(dceba)+(35214)(cebad)+(52143)(ebadc)\\
&+(21354)(baced)+(13542)(acedb)+(35421)(cedba)+(54213)(edbac)+(42135)(dbace)\\
&+(32541)(cbeda)+(25413)(bedac)+(54132)(edacb)+(41325)(dacbe)+(13254)(acbed)\\
&+(35241)(cebda)+(52413)(ebdac)+(24135)(bdace)+(41352)(daceb)+(13524)(acebd)\\
&+(52431)(ebdca)+(24315)(bdcae)+(43152)(dcaeb)+(31524)(caebd)+(15243)(aebdc)\\
&+(42531)(dbeca)+(25314)(becad)+(53142)(ecabd)+(31425)(cabde)+(14253)(abdec)\\
&+(32415)(cbdae)+(24153)(bdaec)+(41532)(daecb)+(15324)(aecbd)+(53241)(ecbda),
\end{split}
\end{equation}
where
\begin{subequations}
\begin{align}
(abc)(def)&\equiv\delta^{f_1}_{h_a}\delta^{f_2}_{h_b}\delta^{f_3}_{h_c}\delta^{i_d}_{j_1}\delta^{i_e}_{j_2}\delta^{i_f}_{j_3},\\
(abcd)(efgh)&\equiv\delta^{f_1}_{h_a}\delta^{f_2}_{h_b}\delta^{f_3}_{h_c}\delta^{f_4}_{h_d}\delta^{i_e}_{j_1}\delta^{i_f}_{j_2}\delta^{i_g}_{j_3}\delta^{i_h}_{j_4},\\
(abcde)(fghij)&\equiv\delta^{f_1}_{h_a}\delta^{f_2}_{h_b}\delta^{f_3}_{h_c}\delta^{f_4}_{h_d}\delta^{f_5}_{h_e}\delta^{i_f}_{j_1}\delta^{i_g}_{j_2}\delta^{i_h}_{j_3}\delta^{i_i}_{j_4}\delta^{i_j}_{j_5}.
\end{align}
\end{subequations}

Other structures are simplified as follows
\begin{equation}
[xyz,lmn]\equiv[lmn]\qquad\textrm{where
}(abc)(def)\equiv\delta^{f_x}_{f_a}\delta^{f_y}_{f_b}\delta^{f_z}_{f_c}\delta^{j_d}_{j_x}\delta^{j_e}_{j_y}\delta^{j_f}_{j_z},
\end{equation}
\begin{equation}
\{ab\}\equiv\delta^{f_a}_{f_8}\delta^{f_b}_{f_{10}}\left(5\delta^{j_8}_{j_a}\delta^{j_{10}}_{j_b}-\delta^{j_8}_{j_b}\delta^{j_{10}}_{j_a}\right)
+\delta^{f_a}_{f_{10}}\delta^{f_b}_{f_8}\left(5\delta^{j_{10}}_{j_a}\delta^{j_8}_{j_b}-\delta^{j_{10}}_{j_b}\delta^{j_8}_{j_a}\right),
\end{equation}
\begin{equation}
\begin{split}
\{abcde\}\equiv\,&\delta^{h_1}_{f_a}\left(\epsilon^{f_bf_ch_2}\epsilon^{f_df_eh_3}+\epsilon^{f_bf_ch_3}\epsilon^{f_df_eh_2}\right)+\delta^{h_2}_{f_a}\left(\epsilon^{f_bf_ch_3}\epsilon^{f_df_eh_1}+\epsilon^{f_bf_ch_1}\epsilon^{f_df_eh_3}\right)\\
&+\delta^{h_3}_{f_a}\left(\epsilon^{f_bf_ch_1}\epsilon^{f_df_eh_2}+\epsilon^{f_bf_ch_2}\epsilon^{f_df_eh_1}\right),
\end{split}
\end{equation}
\begin{equation}
\{abc,de\}\equiv\epsilon^{f_af_bf_c}\epsilon_{j_aj_bj_c}\left[\delta^{f_d}_{f_5}\delta^{f_e}_{f_7}\left(4\delta^{j_5}_{j_d}\delta^{j_7}_{j_e}-\delta^{j_5}_{j_e}\delta^{j_7}_{j_d}\right)
+\delta^{f_d}_{f_7}\delta^{f_e}_{f_5}\left(4\delta^{j_7}_{j_d}\delta^{j_5}_{j_e}-\delta^{j_7}_{j_e}\delta^{j_5}_{j_d}\right)
\right],
\end{equation}
\begin{equation}
\begin{split}
\{abcdef\}\equiv\,&\epsilon^{f_af_bf_c}\epsilon^{f_df_ef_f}\epsilon_{j_aj_bj_c}\epsilon_{j_dj_ej_f}+\epsilon^{f_af_bf_d}\epsilon^{f_cf_ef_f}\epsilon_{j_aj_bj_d}\epsilon_{j_cj_ej_f}+\epsilon^{f_af_bf_e}\epsilon^{f_cf_df_f}\epsilon_{j_aj_bj_e}\epsilon_{j_cj_dj_f}\\
&+\epsilon^{f_af_bf_f}\epsilon^{f_cf_df_e}\epsilon_{j_aj_bj_f}\epsilon_{j_cj_dj_e}+\epsilon^{f_af_cf_d}\epsilon^{f_bf_ef_f}\epsilon_{j_aj_cj_d}\epsilon_{j_bj_ej_f}+\epsilon^{f_af_cf_e}\epsilon^{f_bf_df_f}\epsilon_{j_aj_cj_e}\epsilon_{j_bj_dj_f}\\
&+\epsilon^{f_af_cf_f}\epsilon^{f_bf_df_e}\epsilon_{j_aj_cj_f}\epsilon_{j_bj_dj_e}+\epsilon^{f_af_df_e}\epsilon^{f_bf_cf_f}\epsilon_{j_aj_dj_e}\epsilon_{j_bj_cj_f}+\epsilon^{f_af_df_f}\epsilon^{f_bf_cf_e}\epsilon_{j_aj_dj_f}\epsilon_{j_bj_cj_e}\\
&+\epsilon^{f_af_ef_f}\epsilon^{f_bf_cf_d}\epsilon_{j_aj_ej_f}\epsilon_{j_bj_cj_d},
\end{split}
\end{equation}
\begin{equation}
\{abc,def\}\equiv\epsilon^{f_af_bf_c}\epsilon_{j_aj_bj_c}\left\{7\,[def,579]-2\,([def,597]+[def,975]+[def,759])+([def,795]+[def,957])\right\}.
\end{equation}

\subsection*{A.4 Group integrals and projections onto Fock states}

Spin-flavor wave functions are constructed from the projection of
Fock states onto rotational wave functions. The rotational wave
functions can be found in \cite{Moi}. The $3Q$ state involves three
quarks that are rotated by three $R$ matrices. The $5Q$ state
involves four quarks and one antiquark that are rotated by four $R$
and one $R^\dag$ matrices. So a general $nQ$ state involves
$(n+3)/2$ quarks and $(n-3)/2$ antiquarks that are rotated by
$(n+3)/2$ $R$ and $(n-3)/2$ $R^\dag$ matrices.

\subsection{Projections of the $3Q$ state}

The first integral corresponds to the projection of the $3Q$ state
onto the octet quantum numbers for the $SU(3)$ group
\begin{equation}
\begin{split}
\int\ud
R\,&R^{f_1}_{j_1}R^{f_2}_{j_2}R^{f_3}_{j_3}\left(R^{f_4}_{j_4}R^{\dag
j_5}_{f_5}\right)\\
=&\,\frac{1}{24}\left(\delta^{f_1}_{f_5}\delta^{j_5}_{j_1}\epsilon^{f_2f_3f_4}\epsilon_{j_2j_3j_4}+\delta^{f_2}_{f_5}\delta^{j_5}_{j_2}\epsilon^{f_1f_3f_4}\epsilon_{j_1j_3j_4}
+\delta^{f_3}_{f_5}\delta^{j_5}_{j_3}\epsilon^{f_1f_2f_4}\epsilon_{j_1j_2j_4}+\delta^{f_4}_{f_5}\delta^{j_5}_{j_4}\epsilon^{f_1f_2f_3}\epsilon_{j_1j_2j_3}\right).
\end{split}
\end{equation}
This integral is zero for any other $SU(N)$ group.

The second integral corresponds to the projection of the $3Q$ state
onto the decuplet quantum numbers for any $SU(N)$ group
\begin{equation}
\begin{split}
\int\ud R\,&R^{f_1}_{j_1}R^{f_2}_{j_2}R^{f_3}_{j_3}R_{h_1}^{\dag
i_1}R_{h_2}^{\dag i_2}R_{h_3}^{\dag
i_3}\\=&\,\frac{1}{N(N^2-1)(N^2-4)}\,\{(N^2-2)\,[123]-N\,([213]+[132]+[321])+2\,([231]+[312])\}.
\end{split}
\end{equation}
There is no problem in the case $N=2$ thanks to (\ref{Formule2})
\begin{equation}
\int\ud R\,R^{f_1}_{j_1}R^{f_2}_{j_2}R^{f_3}_{j_3}R_{h_1}^{\dag
i_1}R_{h_2}^{\dag i_2}R_{h_3}^{\dag
i_3}=\frac{1}{24}\,\{3\,[123]-([231]+[312])\}.
\end{equation}

The third integral corresponds to the projection of the antidecuplet
onto the $3Q$ state for the $SU(3)$ group
\begin{equation}
\int\ud
R\,R^{f_1}_{j_1}R^{f_2}_{j_2}R^{f_3}_{j_3}R^{f_4}_{j_4}R^{f_5}_{j_5}R^{f_6}_{j_6}=\frac{1}{72}\,\{123456\}.\label{Three
quarks antidecuplet}
\end{equation}
This integral is also non-vanishing in only two other cases $N=2$
and $N=6$. The (conjugated) rotational wave function of the
antidecuplet is
\begin{equation}
A_k^{*\{h_1h_2h_3\}}(R)=\frac{1}{3}\left(R^{h_1}_3R^{h_2}_3R^{h_3}_k+R^{h_2}_3R^{h_3}_3R^{h_1}_k+R^{h_3}_3R^{h_1}_3R^{h_2}_k\right).\label{Tensor
pentaquarks}
\end{equation}
Due to the antisymmetric structure of (\ref{Three quarks
antidecuplet}) one can see that the projection of the antidecuplet
on the $3Q$ sector is vanishing and thus that pentaquarks cannot be
made of three quarks only.

\subsection{Projections of the $5Q$ state}

The first integral corresponds to the projection of the $5Q$ state
onto the octet quantum numbers for the $SU(3)$ group
\begin{equation}
\begin{split}
\int\ud
R\,&R^{f_1}_{j_1}R^{f_2}_{j_2}R^{f_3}_{j_3}\left(R^{f_4}_{j_4}R^{\dag
j_5}_{f_5}\right)\left(R^{f_6}_{j_6}R^{\dag
j_7}_{f_7}\right)\\=&\,\frac{1}{360}\,[\{123,46\}+\{124,36\}+\{126,34\}+\{134,26\}+\{136,24\}+\{146,23\}\\
&\hspace{6cm}+\{346,12\}+\{246,13\}+\{236,14\}+\{234,16\}].\\
\end{split}
\end{equation}
This integral is zero for any other $SU(N)$ group.

The second integral corresponds to the projection of the $5Q$ state
onto the decuplet quantum numbers for any $SU(N)$ group
\begin{equation}
\begin{split}
\int\ud
R\,&R^{f_1}_{j_1}R^{f_2}_{j_2}R^{f_3}_{j_3}R^{f_4}_{j_4}R_{h_1}^{\dag
i_1}R_{h_2}^{\dag i_2}R_{h_3}^{\dag i_3}R_{h_4}^{\dag
i_4}=\frac{1}{N^2(N^2-1)(N^2-4)(N^2-9)}\\
\times&\,\{(N^4-8N^2+6)\,[1234]-N(N^2-4)\,([2134]+[3214]+[1432]+[1324]+[1243]+[4231])\\
&+(N^2+6)\,([3412]+[2143]+[4321])-5N\,([2341]+[4123]+[3421]+[4312]+[3142]+[2413])\\
&+(2N^2-3)\,([1342]+[4213]+[3241]+[2314]+[3124]+[4132]+[2431]+[1423])\}.
\end{split}
\end{equation}
No problem arises either in the case $N=3$
\begin{equation}
\begin{split}
\int\ud
R\,&R^{f_1}_{j_1}R^{f_2}_{j_2}R^{f_3}_{j_3}R^{f_4}_{j_4}R_{h_1}^{\dag
i_1}R_{h_2}^{\dag i_2}R_{h_3}^{\dag i_3}R_{h_4}^{\dag
i_4}\\
=&\,\frac{1}{2160}\,\{48\,[1234]-11\,([2134]+[3214]+[1432]+[1324]+[1243]+[4231])\\
&-6\,([3412]+[2143]+[4321])+7\,([2341]+[4123]+[3421]+[4312]+[3142]+[2413])\}.
\end{split}
\end{equation}
or in the case $N=2$ thanks to the generalization of
(\ref{Formule2})
\begin{equation}
\begin{split}
\int\ud
R\,&R^{f_1}_{j_1}R^{f_2}_{j_2}R^{f_3}_{j_3}R^{f_4}_{j_4}R_{h_1}^{\dag
i_1}R_{h_2}^{\dag i_2}R_{h_3}^{\dag i_3}R_{h_4}^{\dag
i_4}\\
=&\,\frac{1}{240}\,\{8\,[1234]-3\,([2341]+[4123]+[3421]+[4312]+[3142]+[2413])\\
&\hspace{8cm}+4\,([3412]+[2143]+[4321])\}.
\end{split}
\end{equation}

The third integral corresponds to the projection of the $5Q$ state
onto the antidecuplet quantum numbers for the $SU(3)$ group
\begin{equation}
\begin{split}
\int\ud
R\,&R^{f_1}_{j_1}R^{f_2}_{j_2}R^{f_3}_{j_3}R^{f_4}_{j_4}R^{f_5}_{j_5}R^{f_6}_{j_6}\left(R^{f_7}_{j_7}R^{\dag
j_8}_{f_8}\right)\\
=&\,\frac{1}{360}\left[\delta^{f_1}_{f_8}\delta^{j_8}_{j_1}\{234567\}+\delta^{f_2}_{f_8}\delta^{j_8}_{j_2}\{134567\}+\delta^{f_3}_{f_8}\delta^{j_8}_{j_3}\{124567\}+\delta^{f_4}_{f_8}\delta^{j_8}_{j_4}\{123567\}\right.\\
&\hspace{5cm}+\delta^{f_5}_{f_8}\delta^{j_8}_{j_5}\{123467\}+\delta^{f_6}_{f_8}\delta^{j_8}_{j_6}\{123457\}+\delta^{f_7}_{f_8}\delta^{j_8}_{j_7}\{123456\}\Big].\\
\end{split}
\end{equation}
This integral is also non-vanishing in only two other cases $N=2$
and $N=6$. The (conjugated) rotational wave function of the
antidecuplet (\ref{Tensor pentaquarks}) is symmetric with respect to
three flavor indices $h_1,h_2,h_3$. The projection of the $5Q$ state
is thus reduced to
\begin{equation}
\begin{split}
\int\ud
R\,&R^{f_1}_{j_1}R^{f_2}_{j_2}R^{f_3}_{j_3}\left(R^{f_4}_{j_4}R^{\dag
j_5}_{f_5}\right)A_k^{*\{h_1h_2h_3\}}(R)\\
=&\,\frac{1}{1080}\Big\{\{51234\}\left(\delta^{j_5}_k\epsilon_{j_1j_23}\epsilon_{j_3j_43}+\delta^{j_5}_3\epsilon_{j_1j_2k}\epsilon_{j_3j_43}+\delta^{j_5}_3\epsilon_{j_1j_23}\epsilon_{j_3j_4k}\right)\\
&+\{52341\}\left(\delta^{j_5}_k\epsilon_{j_2j_33}\epsilon_{j_4j_13}+\delta^{j_5}_3\epsilon_{j_2j_3k}\epsilon_{j_4j_13}+\delta^{j_5}_3\epsilon_{j_2j_33}\epsilon_{j_4j_1k}\right)\\
&+\{51324\}\left(\delta^{j_5}_k\epsilon_{j_1j_33}\epsilon_{j_2j_43}+\delta^{j_5}_3\epsilon_{j_1j_3k}\epsilon_{j_2j_43}+\delta^{j_5}_3\epsilon_{j_1j_33}\epsilon_{j_2j_4k}\right)\Big\}.
\end{split}
\end{equation}

\subsection{Projections of the $7Q$ state}

The first integral corresponds to the projection of the $7Q$ state
onto the octet quantum numbers for the $SU(3)$ group
\begin{equation}
\begin{split}
\int\ud
R\,&R^{f_1}_{j_1}R^{f_2}_{j_2}R^{f_3}_{j_3}\left(R^{f_4}_{j_4}R^{\dag
j_5}_{f_5}\right)\left(R^{f_6}_{j_6}R^{\dag
j_7}_{f_7}\right)\left(R^{f_8}_{j_8}R^{\dag
j_9}_{f_9}\right)\\
=&\,\frac{1}{2160}\,(\{123,468\}+\{124,368\}+\{126,348\}+\{128,346\}+\{134,268\}+\{136,248\}+\{138,246\}\\
&+\{146,238\}+\{148,236\}+\{168,234\}+\{468,123\}+\{368,124\}+\{348,126\}+\{346,128\}\\
&+\{268,134\}+\{248,136\}+\{246,138\}+\{238,146\}+\{236,148\}+\{234,168\}).
\end{split}
\end{equation}
This integral is zero for any other $SU(N)$ group.

The second integral corresponds to the projection of the $7Q$ state
onto the decuplet quantum numbers for any $SU(N)$ group
\begin{equation}
\begin{split}
\int\ud
R\,&R^{f_1}_{j_1}R^{f_2}_{j_2}R^{f_3}_{j_3}R^{f_4}_{j_4}R^{f_5}_{j_5}R_{h_1}^{\dag
i_1}R_{h_2}^{\dag i_2}R_{h_3}^{\dag i_3}R_{h_4}^{\dag
i_4}R_{h_5}^{\dag
i_5}\\
=&\,\frac{1}{N^2(N^2-1)(N^2-4)(N^2-9)(N^2-16)}\,\{N(N^4-20N^2+78)\,[12345]\\
&-(N^4-14N^2+24)([21345]+[52341]+[12354]+[12435]+[13245]+[14325]+[32145]\\
&\quad+[15342]+[42315]+[12543])\\
&-2(N^2+12)\,([34521]+[34152]+[35412]+[43512]+[24513]+[54123]+[35124]+[45132]\\
&\quad+[45213]+[41523]+[21534]+[54231]+[31254]+[51432]+[53214]+[25431]+[43251]\\
&\quad+[21453]+[53421]+[23154])\\
&+2N(N^2-9)\,([12453]+[23145]+[42351]+[15324]+[15243]+[32415]+[24315]+[14352]\\
&\quad+[14235]+[51342]+[52314]+[13425]+[25341]+[52143]+[42135]+[41325]+[13542]\\
&\quad+[32541]+[12534]+[31245])\\
&+N(N^2-2)\,([54321]+[32154]+[15432]+[43215]+[21543]+[45312]+[42513]+[14523]\\
&\quad+[34125]+[35142]+[21354]+[52431]+[13254]+[21435]+[53241])\\
&+14N\,([23451]+[31452]+[53412]+[23514]+[24531]+[34251]+[41253]+[51423]\\
&\quad+[53124]+[25134]+[45231]+[51234]+[25413]+[43521]+[24153]+[35421]+[43152]\\
&\quad+[41532]+[54213]+[31524]+[54132]+[35214]+[45123]+[34512])\\
&-(5N^2-24)\,([13452]+[23415]+[23541]+[24351]+[32451]+[41352]+[52413]+[13524]\\
&\quad+[24135]+[35241]+[53142]+[25314]+[42531]+[14253]+[31425]+[15234]+[41235]\\
&\quad+[51243]+[51324]+[52134]+[15423]+[43125]+[25143]+[45321]+[42153]+[14532]\\
&\quad+[34215]+[31542]+[54312]+[32514])\}.
\end{split}
\end{equation}
No problem arises in the case $N=4$
\begin{equation}
\begin{split}
\int\ud
R\,&R^{f_1}_{j_1}R^{f_2}_{j_2}R^{f_3}_{j_3}R^{f_4}_{j_4}R^{f_5}_{j_5}R_{h_1}^{\dag
i_1}R_{h_2}^{\dag i_2}R_{h_3}^{\dag i_3}R_{h_4}^{\dag
i_4}R_{h_5}^{\dag
i_5}=\frac{1}{80640}\,\{179\,[12345]\\
&-52\,([21345]+[52341]+[12354]+[12435]+[13245]+[14325]+[32145]+[15342]+[42315]\\
&\quad+[12543])\\
&+12\,([34521]+[34152]+[35412]+[43512]+[24513]+[54123]+[35124]+[45132]+[45213]\\
&\quad+[41523]+[21534]+[54231]+[31254]+[51432]+[53214]+[25431]+[43251]+[21453]\\
&\quad+[53421]+[23154])\\
&+19\,([12453]+[23145]+[42351]+[15324]+[15243]+[32415]+[24315]+[14352]+[14235]\\
&\quad+[51342]+[52314]+[13425]+[25341]+[52143]+[42135]+[41325]+[13542]+[32541]\\
&\quad+[12534]+[31245])\\
&+3\,([54321]+[32154]+[15432]+[43215]+[21543]+[45312]+[42513]+[14523]+[34125]\\
&\quad+[35142]+[21354]+[52431]+[13254]+[21435]+[53241])\\
&-13\,([23451]+[31452]+[53412]+[23514]+[24531]+[34251]+[41253]+[51423]+[53124]\\
&\quad+[25134]+[45231]+[51234]+[25413]+[43521]+[24153]+[35421]+[43152]+[41532]\\
&\quad+[54213]+[31524]+[54132]+[35214]+[45123]+[34512])\},
\end{split}
\end{equation}
in the case $N=3$
\begin{equation}
\begin{split}
\int\ud
R\,&R^{f_1}_{j_1}R^{f_2}_{j_2}R^{f_3}_{j_3}R^{f_4}_{j_4}R^{f_5}_{j_5}R_{h_1}^{\dag
i_1}R_{h_2}^{\dag i_2}R_{h_3}^{\dag i_3}R_{h_4}^{\dag
i_4}R_{h_5}^{\dag
i_5}=\frac{1}{15120}\,\{151\,[12345]\\
&-38\,([21345]+[52341]+[12354]+[12435]+[13245]+[14325]+[32145]+[15342]+[42315]\\
&\quad+[12543])\\
&-2\,([34521]+[34152]+[35412]+[43512]+[24513]+[54123]+[35124]+[45132]+[45213]\\
&\quad+[41523]+[21534]+[54231]+[31254]+[51432]+[53214]+[25431]+[43251]+[21453]\\
&\quad+[53421]+[23154])\\
&+10\,([12453]+[23145]+[42351]+[15324]+[15243]+[32415]+[24315]+[14352]+[14235]\\
&\quad+[51342]+[52314]+[13425]+[25341]+[52143]+[42135]+[41325]+[13542]+[32541]\\
&\quad+[12534]+[31245])\\
&+5\,([54321]+[32154]+[15432]+[43215]+[21543]+[45312]+[42513]+[14523]+[34125]\\
&\quad+[35142]+[21354]+[52431]+[13254]+[21435]+[53241])\}
\end{split}
\end{equation}
or in the case $N=2$ thanks to the generalization of
(\ref{Formule2})
\begin{equation}
\begin{split}
\int\ud
R\,&R^{f_1}_{j_1}R^{f_2}_{j_2}R^{f_3}_{j_3}R^{f_4}_{j_4}R^{f_5}_{j_5}R_{h_1}^{\dag
i_1}R_{h_2}^{\dag i_2}R_{h_3}^{\dag i_3}R_{h_4}^{\dag
i_4}R_{h_5}^{\dag
i_5}=\frac{1}{1440}\,\{57\,[12345]\\
&-11\,([21345]+[52341]+[12354]+[12435]+[13245]+[14325]+[32145]+[15342]+[42315]\\
&\quad+[12543])\\
&+2\,([12453]+[23145]+[42351]+[15324]+[15243]+[32415]+[24315]+[14352]+[14235]\\
&\quad+[51342]+[52314]+[13425]+[25341]+[52143]+[42135]+[41325]+[13542]+[32541]\\
&\quad+[12534]+[31245])\\
&+([54321]+[32154]+[15432]+[43215]+[21543]+[45312]+[42513]+[14523]+[34125]\\
&\quad+[35142]+[21354]+[52431]+[13254]+[21435]+[53241])\}.
\end{split}
\end{equation}
\newpage

The third integral corresponds to the projection of the $7Q$ state
onto the antidecuplet quantum numbers for the $SU(3)$ group
\begin{equation}
\begin{split}
\int\ud
R\,&R^{f_1}_{j_1}R^{f_2}_{j_2}R^{f_3}_{j_3}R^{f_4}_{j_4}R^{f_5}_{j_5}R^{f_6}_{j_6}\left(R^{f_7}_{j_7}R^{\dag
j_8}_{f_8}\right)\left(R^{f_9}_{j_9}R^{\dag
j_{10}}_{f_{10}}\right)\\
=&\,\frac{1}{8640}\,[\{123456\}\{79\}+\{123457\}\{69\}+\{123467\}\{59\}+\{123567\}\{49\}+\{124567\}\{39\}\\
&+\{134567\}\{29\}+\{234567\}\{19\}+\{123459\}\{67\}+\{123469\}\{57\}+\{123569\}\{47\}\\
&+\{124569\}\{37\}+\{134569\}\{27\}+\{234569\}\{17\}+\{123479\}\{56\}+\{123579\}\{46\}\\
&+\{124579\}\{36\}+\{134579\}\{26\}+\{234579\}\{16\}+\{123679\}\{45\}+\{124679\}\{35\}\\
&+\{134679\}\{25\}+\{234679\}\{15\}+\{125679\}\{34\}+\{135679\}\{24\}+\{235679\}\{14\}\\
&+\{145679\}\{23\}+\{245679\}\{13\}+\{345679\}\{12\}].
\end{split}
\end{equation}
This integral is also non-vanishing in only two other cases $N=2$
and $N=6$. The (conjugated) rotational wave function of the
antidecuplet (\ref{Tensor pentaquarks}) is symmetric with respect to
three flavor indices $h_1,h_2,h_3$. The projection onto the $7Q$
state is thus reduced to
\begin{equation}
\begin{split}
\int\ud
R\,&R^{f_1}_{j_1}R^{f_2}_{j_2}R^{f_3}_{j_3}\left(R^{f_4}_{j_4}R^{\dag
j_5}_{f_5}\right)\left(R_{j_6}^{\dag f_6}R_{f_7}^{\dag
j_7}\right)A_k^{*\{h_1h_2h_3\}}(R)\\
=&\,\frac{1}{25920}\,\Big\{\left[\delta^{f_1}_{f_5}\{72346\}\left(5\delta^{j_5}_{j_1}\delta^{j_7}_k-\delta^{j_7}_{j_1}\delta^{j_5}_k\right)+\delta^{f_1}_{f_7}\{52346\}\left(5\delta^{j_7}_{j_1}\delta^{j_5}_k-\delta^{j_5}_{j_1}\delta^{j_7}_k\right)\right]\\
&+\left[\delta^{f_1}_{f_5}\{72436\}\left(5\delta^{j_5}_{j_1}\delta^{j_7}_k-\delta^{j_7}_{j_1}\delta^{j_5}_k\right)+\delta^{f_1}_{f_7}\{52436\}\left(5\delta^{j_7}_{j_1}\delta^{j_5}_k-\delta^{j_5}_{j_1}\delta^{j_7}_k\right)\right]\\
&+\left[\delta^{f_1}_{f_5}\{72634\}\left(5\delta^{j_5}_{j_1}\delta^{j_7}_k-\delta^{j_7}_{j_1}\delta^{j_5}_k\right)+\delta^{f_1}_{f_7}\{52634\}\left(5\delta^{j_7}_{j_1}\delta^{j_5}_k-\delta^{j_5}_{j_1}\delta^{j_7}_k\right)\right]\\
&+\left[\delta^{f_2}_{f_5}\{71346\}\left(5\delta^{j_5}_{j_2}\delta^{j_7}_k-\delta^{j_7}_{j_2}\delta^{j_5}_k\right)+\delta^{f_2}_{f_7}\{51346\}\left(5\delta^{j_7}_{j_2}\delta^{j_5}_k-\delta^{j_5}_{j_2}\delta^{j_7}_k\right)\right]\\
&+\left[\delta^{f_2}_{f_5}\{71436\}\left(5\delta^{j_5}_{j_2}\delta^{j_7}_k-\delta^{j_7}_{j_2}\delta^{j_5}_k\right)+\delta^{f_2}_{f_7}\{51436\}\left(5\delta^{j_7}_{j_2}\delta^{j_5}_k-\delta^{j_5}_{j_2}\delta^{j_7}_k\right)\right]\\
&+\left[\delta^{f_2}_{f_5}\{71634\}\left(5\delta^{j_5}_{j_2}\delta^{j_7}_k-\delta^{j_7}_{j_2}\delta^{j_5}_k\right)+\delta^{f_2}_{f_7}\{51634\}\left(5\delta^{j_7}_{j_2}\delta^{j_5}_k-\delta^{j_5}_{j_2}\delta^{j_7}_k\right)\right]\\
&+\left[\delta^{f_3}_{f_5}\{71246\}\left(5\delta^{j_5}_{j_3}\delta^{j_7}_k-\delta^{j_7}_{j_3}\delta^{j_5}_k\right)+\delta^{f_3}_{f_7}\{51246\}\left(5\delta^{j_7}_{j_3}\delta^{j_5}_k-\delta^{j_5}_{j_3}\delta^{j_7}_k\right)\right]\\
&+\left[\delta^{f_3}_{f_5}\{71426\}\left(5\delta^{j_5}_{j_3}\delta^{j_7}_k-\delta^{j_7}_{j_3}\delta^{j_5}_k\right)+\delta^{f_3}_{f_7}\{51426\}\left(5\delta^{j_7}_{j_3}\delta^{j_5}_k-\delta^{j_5}_{j_3}\delta^{j_7}_k\right)\right]\\
&+\left[\delta^{f_3}_{f_5}\{71624\}\left(5\delta^{j_5}_{j_3}\delta^{j_7}_k-\delta^{j_7}_{j_3}\delta^{j_5}_k\right)+\delta^{f_3}_{f_7}\{51624\}\left(5\delta^{j_7}_{j_3}\delta^{j_5}_k-\delta^{j_5}_{j_3}\delta^{j_7}_k\right)\right]\\
&+\left[\delta^{f_4}_{f_5}\{71236\}\left(5\delta^{j_5}_{j_4}\delta^{j_7}_k-\delta^{j_7}_{j_4}\delta^{j_5}_k\right)+\delta^{f_4}_{f_7}\{51236\}\left(5\delta^{j_7}_{j_4}\delta^{j_5}_k-\delta^{j_5}_{j_4}\delta^{j_7}_k\right)\right]\\
&+\left[\delta^{f_4}_{f_5}\{71326\}\left(5\delta^{j_5}_{j_4}\delta^{j_7}_k-\delta^{j_7}_{j_4}\delta^{j_5}_k\right)+\delta^{f_4}_{f_7}\{51326\}\left(5\delta^{j_7}_{j_4}\delta^{j_5}_k-\delta^{j_5}_{j_4}\delta^{j_7}_k\right)\right]\\
&+\left[\delta^{f_4}_{f_5}\{71623\}\left(5\delta^{j_5}_{j_4}\delta^{j_7}_k-\delta^{j_7}_{j_4}\delta^{j_5}_k\right)+\delta^{f_4}_{f_7}\{51623\}\left(5\delta^{j_7}_{j_4}\delta^{j_5}_k-\delta^{j_5}_{j_4}\delta^{j_7}_k\right)\right]\\
&+\left[\delta^{f_6}_{f_5}\{71234\}\left(5\delta^{j_5}_{j_6}\delta^{j_7}_k-\delta^{j_7}_{j_6}\delta^{j_5}_k\right)+\delta^{f_6}_{f_7}\{51234\}\left(5\delta^{j_7}_{j_6}\delta^{j_5}_k-\delta^{j_5}_{j_6}\delta^{j_7}_k\right)\right]\\
&+\left[\delta^{f_6}_{f_5}\{71324\}\left(5\delta^{j_5}_{j_6}\delta^{j_7}_k-\delta^{j_7}_{j_6}\delta^{j_5}_k\right)+\delta^{f_6}_{f_7}\{51324\}\left(5\delta^{j_7}_{j_6}\delta^{j_5}_k-\delta^{j_5}_{j_6}\delta^{j_7}_k\right)\right]\\
&+\left[\delta^{f_6}_{f_5}\{71423\}\left(5\delta^{j_5}_{j_6}\delta^{j_7}_k-\delta^{j_7}_{j_6}\delta^{j_5}_k\right)+\delta^{f_6}_{f_7}\{51423\}\left(5\delta^{j_7}_{j_6}\delta^{j_5}_k-\delta^{j_5}_{j_6}\delta^{j_7}_k\right)\right]\Big\}.
\end{split}
\end{equation}

\end{document}